\documentclass[useAMS]{mn2e}
\usepackage{graphicx}
\usepackage[T1]{fontenc}
\usepackage{aecompl}
\def\simlt{\lower.5ex\hbox{$\; \buildrel < \over \sim \;$}}
\def\simgt{\lower.5ex\hbox{$\; \buildrel > \over \sim \;$}}
\usepackage{caption}
\DeclareCaptionFormat{cont}{#1 (cont.)#2#3\par}

\title[KAT--7 observations of a mass-selected sample of galaxy clusters]
{KAT--7 observations of an unbiased sample of mass-selected galaxy clusters}

\author[G. Bernardi et al.]
  {G.~Bernardi$^{1,2,3}$\thanks{E-mail: gbernardi@ska.ac.za},
  T.~Venturi$^{4
  }$, R.~Cassano$^4$, D.~Dallacasa$^5$, G.~Brunetti$^4$, V. Cuciti$^{4,5}$, \newauthor
  M.~Johnston--Hollitt$^6$, N.~Oozeer$^{1,7,8}$, V. Parekh$^9$ and O.M.~Smirnov$^{2,1}$\\
  $^1$SKA SA, 3rd Floor, The Park, Park Road, Pinelands, 7405, South Africa\\
  $^2$Department of Physics and Electronics, Rhodes University, PO Box 94, Grahamstown, 6140, South Africa\\
  $^3$Harvard-Smithsonian Center for Astrophysics, Garden Street 60, Cambridge, MA, 02138\\
  $^4$INAF - Istituto di Radioastronomia, via Gobetti 101, 40129 Bologna, Italy\\
  $^5$Dipartimento di Fisica e Astronomia, Universit\'a di Bologna, viale Berti Pichat 6/2, I-40127 Bologna, Italy\\
  $^6$School of Chemical \& Physical Sciences, Victoria University of Wellington, Wellington 6140, New Zealand\\
  $^7$African Institute for Mathematical Sciences, 6-8 Melrose Road, Muizenberg 7945, South Africa\\
  $^8$Centre for Space Research, North-West University, Potchefstroom 2520, South Africa\\  
  $^9$Raman Research Institute, C.V. Raman Avenue, Sadashivanagar, Near Mekhri circle, Bengaluru, Karnataka 560080, India}

\begin{document}

\date{Accepted xxxx. Received yyyy; in original form zzzz}

\pagerange{\pageref{firstpage}--\pageref{lastpage}} \pubyear{2009}

\maketitle

\label{firstpage}

\begin{abstract}
The presence of megaparsec-scale radio halos in galaxy clusters has already 
been established by many observations over the last two decades. The emerging 
explanation for the formation of these giant sources of diffuse synchrotron
radio emission is that they trace turbulent regions in the intracluster 
medium, where particles are trapped and accelerated during cluster mergers. Our current observational knowledge is, however, mainly limited to massive systems.
Here we present observations of a sample of 14 mass-selected galaxy clusters,
i.e. $M_{\rm 500} > 4\times10^{14}$~M${_\odot}$, in the Southern 
Hemisphere, aimed to study the occurrence of radio halos in low mass 
clusters and test the correlation between the radio halo power at 1.4 GHz 
$P_{\rm 1.4}$ and the cluster mass $M_{\rm 500}$. Our observations were
performed with the 7-element Karoo Array Telescope at 1.86 GHz.
We found three candidates to host diffuse cluster-scale emission 
and derived upper limits at the level of 
$0.6 - 1.9 \times 10^{24}$~Watt~Hz$^{-1}$ for $\sim 50\%$ of the clusters in 
the sample, significantly increasing the number of clusters with radio halo information in the considered mass range. Our results confirm that bright radio halos in less massive galaxy clusters are statistically rare.
\end{abstract}


\vskip 3cm

\section{Introduction}
\label{intro:Sect}
In the hierarchical model of structure formation, galaxy clusters form through 
the merging of smaller substructures. A fraction of the energy dissipated 
during these merging events can be channelled into the amplification of 
magnetic fields and acceleration of relativistic particles (see Brunetti \& Jones 2014 for a review on the topic). Giant radio 
halos (RHs) are potential probes of this process. They are Mpc-scale 
diffuse radio sources with steep spectrum and low surface brightness that are 
found in the central regions of a number of galaxy clusters 
(see Feretti et al. 2012 for a recent observational review).
In this respect clusters are crossroads between cosmology and 
plasma astrophysics.  

Giant RHs have no obvious optical counterpart, and rather show a remarkable 
connection with the intracluster medium: their extent and morphology are 
closely related to the thermal Bremsstrahlung X--ray emission.
One recent observational milestone, based on the Extended GMRT Radio Halo 
Cluster Survey (EGRHS, Venturi et al. 2007, 2008, Kale et al. 2013, 2015) and 
complemented with results from the literature, is that RHs are hosted 
in only $\sim$~20--30\% of X--ray luminous clusters
($L_{\rm X} > 5 \times 10^{44}$ erg~s$^{-1}$).
Furthermore, clusters branch into two populations in the 
$\log{L_{\rm X}} - \log{P_{\rm 1.4}}$ plane: they either host a RH 
whose radio power shows a tight correlation with the cluster X--ray luminosity $L_{\rm X}$,
or are ``radio--quiet'', with upper limits to the radio power of the 
undetected RH well below the correlation (Brunetti et al. 2007).

Remarkably, there is a tight connection between these RHs and cluster 
mergers (Cassano et al. 2010, 2013):
a detailed and quantitative radio/X--ray analysis of the clusters in the EGRHS 
shows that giant radio halos are found only in dynamically interacting 
(merging) systems, whereas relaxed clusters do not produce diffuse emission 
on the Mpc-scale at the sensitivity level of current observations.
 
The recent advent of cluster surveys through the detection of the 
Sunyaev--Zeldovich (SZ) effect (e.g. with the Planck satellite) has allowed the study of the 
$P_{\rm 1.4}$--$M_{\rm 500}$\footnote{where $M_{500}$ is the total cluster mass 
within the radius $R_{500}$, defined as the radius corresponding to a total 
density contrast $500\rho_c(z)$, where $\rho_c(z)$ is the critical density 
of the Universe at  the cluster redshift.} correlation (Basu et al. 2012, Cassano et al. 2013) , which shows that massive ($M_{500} > 5.5\times10^{14}$ M$_{\odot}$) clusters have a bimodal behaviour also in the radio--SZ diagram.

The commonly adopted scenario to explain the origin of RHs and their 
connection with mergers is based on the (re)acceleration of relativistic 
electrons by turbulence generated during clusters mergers (i.e., Brunetti \& Jones 2014).
According to this scenario the formation and evolution of RHs depends on the 
cluster merging rate at the different cosmic epochs and on the mass of the 
hosting clusters, which ultimately sets the energy budget that can be 
dissipated and drained in the relativistic particles and magnetic fields. 
One of the key expectations is that giant RHs should be found in 
massive/energetic merger events, and become rarer in less massive clusters 
and eventually absent in relaxed systems (Cassano \& Brunetti 2005).
\\
In line with this scenario, Cuciti et al. (2015) recently found some evidences for an increasing occurrence of radio halos with the cluster mass. However, present statistical studies are essentially limited to very X-ray luminous ($L_{X [0.1-2.4 \, {\rm KeV}]} > 5 \times 10^{44}$~erg~s$^{-1}$) and massive ($M_{500} \ge 6 \times 10^{14} \, {\rm M}_{\odot}$) galaxy clusters.
Extending these constraints both in redshift and to lower 
cluster masses is vital to understand the origin of cluster-scale radio 
emission and to test the predictions of theoretical models. However, this 
means to explore the faint end of the correlation between cluster mass (or 
X--ray luminosity) and radio power, which is challenging from an observational 
point of view as RHs are expected to become progressively fainter in less massive systems.

A new opportunity is offered by the improved imaging sensitivity and sampling 
of angular scales of the new generation of radio interferometers.  
In particular, the precursors of the Square Kilometre Array (SKA) such as 
ASKAP and MeerKAT in the GHz regime and LOFAR (van Haarlem et al. 2013) and 
MWA (Tingay et al. 2013) at frequencies below 300~MHz, allow to extend 
earlier studies to a preliminary exploration of low mass nearby clusters.

Here we present the result of a pilot observational study carried out with the 7-antenna Karoo Array Telescope (KAT--7) designed to extend our knowledge 
of the occurrence of RHs in galaxy clusters to a mass--limited sample 
($M_{\rm 500} > 4 \times 10^{14}$ M$_{\odot}$) of nearby clusters in the Southern 
Hemisphere. 

The paper is organized as follows: in Section~\ref{sec:sample} we present
the sample of galaxy clusters selected for this study; in Section 3
we describe the observations and the data reduction; in 
Section 4 we discuss the $\log{P_{\rm 1.4}}$ -- $\log{M_{\rm 500}}$  correlation in the 
light of our results, and we conclude in 
Section 5.
%
%
\begin{figure}
\centering
\includegraphics[width=1\columnwidth]{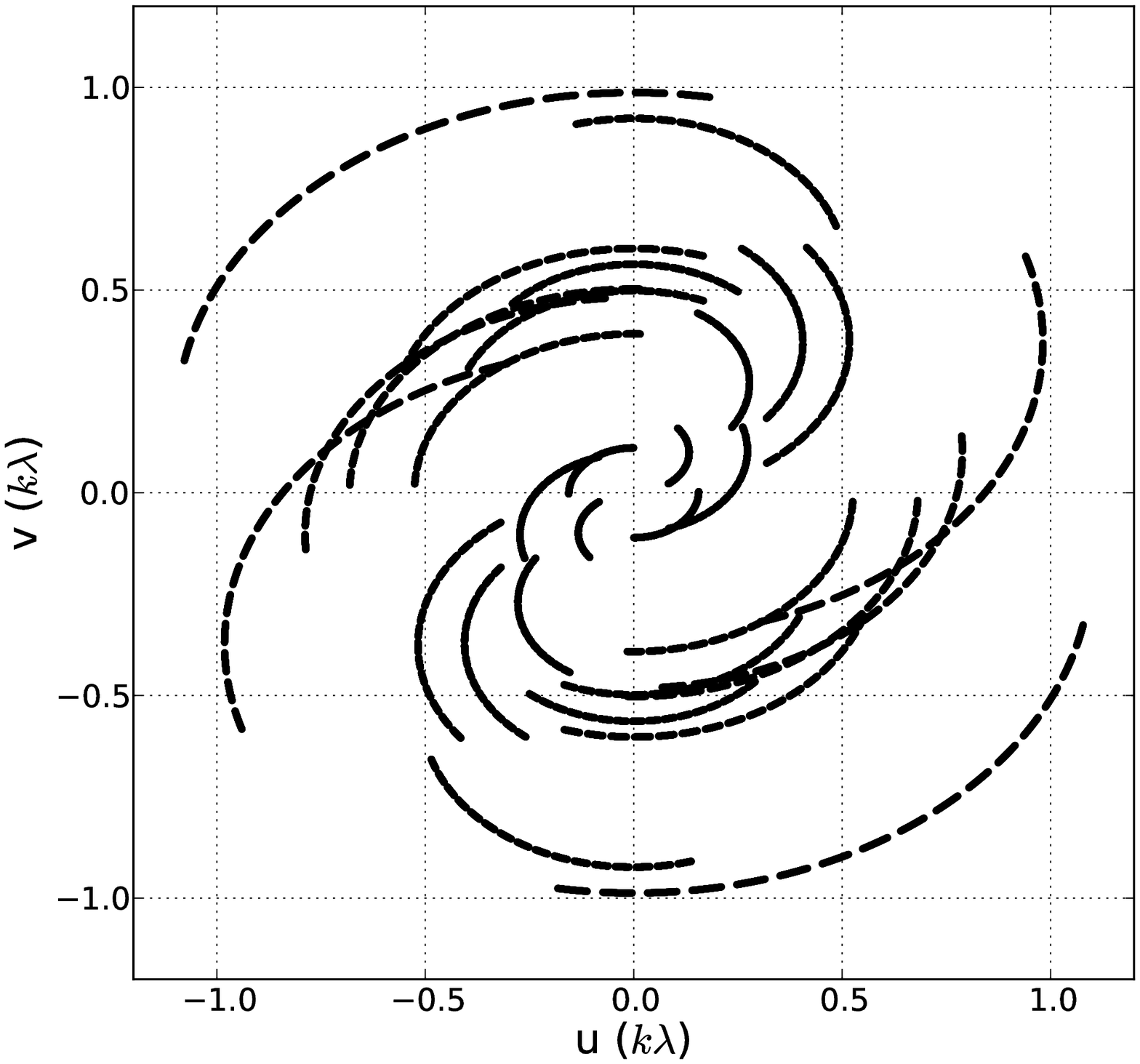}
\includegraphics[width=1\columnwidth]{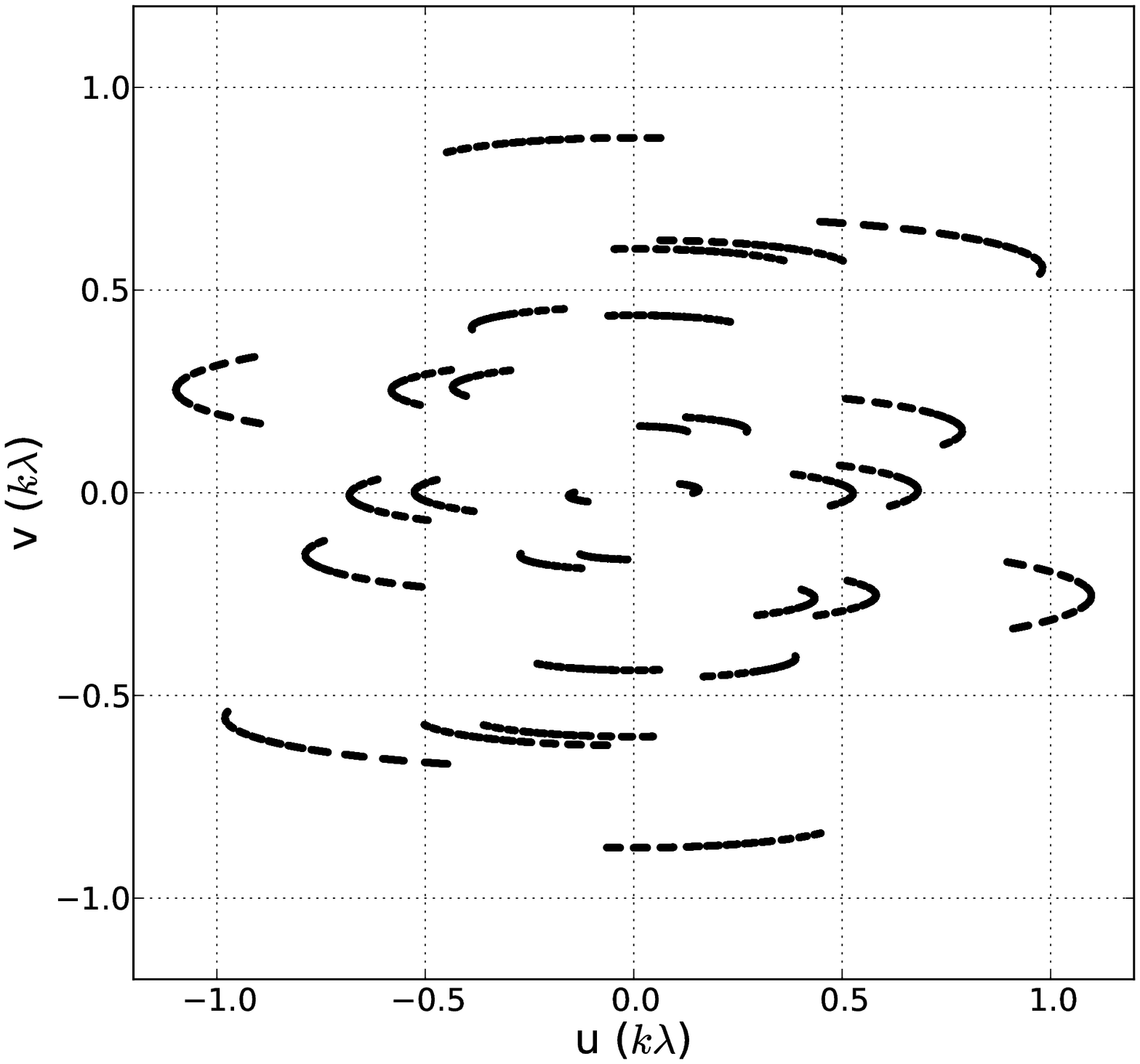}
\caption{Monochromatic $uv$~coverage for RXCJ\,1358.9--4750 (top) and 
A\,644 (bottom).}
\label{fig:uv-coverage}
\end{figure}
%

\section{The cluster sample}
\label{sec:sample}

In order to complement the results based on the EGRHS, and to increase the 
information on the presence of diffuse radio emission (in the form of giant 
RHs) to lower mass clusters in the nearby Universe, we selected all 
the galaxy clusters from the Planck SZ Cluster Catalogue (Planck Collaboration, 2014) with the following criteria:

\begin{itemize}
\item[] mass, $M_{\rm 500} > 4 \times 10^{14}$ M$_{\odot}$;
\item[] redshift range, $0.05 < {\rm z} < 0.11$\footnote{Note that a linear 
size of 1 Mpc corresponds to $\sim17$~arcmin at z~$= 0.05$ and $\sim9$~arcmin at z~$= 0.1$.}; 
\item[] $\delta < 0^{\circ}$.
\end{itemize}
The sample includes 21 clusters, listed in Table~\ref{tab:logs} 
in order of decreasing mass. Except for RXC~J\,1407.8--5100, all clusters 
have high quality X--ray images, from $Chandra$ and/or {\it XMM--Newton}. 
For four of them the presence of diffuse radio emission is known from the 
literature, hence they were not observed. In particular,
RXC~J\,1638.2--6420 (the Triangulum Australis, Scaife et al., 2015) and 
A\,754  (Macario et al. 2011) host a RH,  
A\,3667 (R\"ottgering et al. 1997) is well--known for the double relics,
and A\,85 (Slee et al. 2001) hosts a diffuse steep--spectrum relic. 
Only 843 MHz images with a 43~arcsec~$\times$~43~arcsec angular resolution from the Sydney University Molonglo Sky Survey (SUMSS, Bock, Large \& Sadler 1999) or 1.4 GHz images at the resolution of 45~arcsec~$\times$~45~arcsec from the Northern VLA Sky Survey (NVSS, Condon et al., 1998) are available for the remaining 17 clusters, which were observed with KAT--7.

\section{Observations and data reduction}
\label{sec:obs}

Our observations were carried out in February and March 2014 with KAT--7, the MeerKAT precursor array that comprises seven antennas, each with a diameter of 12~m, distributed in a randomized configuration that maximizes the $uv$ coverage in already $\sim$~4~hours, with baseline lengths ranging from 26~m to 185~m. It is located near the SKA core site in the Karoo semi-desert area, about 80 km north-west of Carnarvon in the Northern Cape, South Africa. KAT--7 operates in the $1.2-1.95$~GHz frequency range with an instantaneous 256~MHz bandwith correlated for continuum observations. Our observations were conducted at the central frequency of 1.86~GHz, with an effective (i.e. after excision of bad channels) 156.64~MHz bandwidth and a 390.625~kHz channel width. As the Triangulum Australis cluster was already observed, we included it in our analysis by using archive data taken at the central frequency of 1.33~GHz. 

Out of the 17 clusters selected, A\,1651 was not observed, data 
for A\,2420 were contaminated by the Sun at the edge of the primary beam and 
A\,2384 was only observed for three hours due to technical problems, giving a 
very limited $uv$ coverage compared to the remaining targets (see 
Table~\ref{tab:obs}). 

The cluster sample presented in this paper therefore contains 14 targets (although our analysis of the $P_{\rm 1.4} - M_{\rm 500}$ correlation will include A\,754 and the Triangulum Australis clusters for which data were already available), each of them observed for 5~hours minimum in order to obtain a good $uv$ coverage.
Typical $uv$~coverages are shown in Figure~\ref{fig:uv-coverage} 
for a high elevation (RXCJ\,1358.9--4750, 
DEC$_{\rm J2000}=-47^{\circ}50^{\prime}49^{\prime\prime}$) and for a low elevation 
(A\,644, DEC$_{\rm J2000}=-07^{\circ}30^{\prime}46^{\prime\prime}$) cluster respectively.
Observation details are reported in Table~\ref{tab:obs}. 

The source PKS\,1934--68 was used as bandpass and flux calibrator for each 
target and was observed several times throughout each observing run. 
PKS\,1934--68 was modelled as a point source with flux density of 13.16~Jy at 
the central observing frequency according to the 
Perley \& Butler (2013) flux scale.

%
\begin{table*}
\caption[]{The low--mass cluster sample. Clusters are listed in order of 
decreasing $M_{500}$ values. The coordinates are referred to the X--ray 
cluster centre.}
\begin{center}
\begin{tabular}{llcccccc}
\hline
\hline\noalign{\smallskip}
Cluster name & Other Name & RA$_{\rm J2000}$ & DEC$_{\rm J2000}$ &   z  & kpc/'' &  M$_{\rm 500}$           & Ref. \\
        &            &                  &                   &         &
& (10$^{14}$M$_{\odot}$)  &      \\
\hline\noalign{\smallskip}
RXCJ\,1638.2--6420  & Triangulum  & 16$^{\rm h}$ 38$^{\rm m}$ 18$^{\rm s}$.3  & --64$^{\circ}$ 21$^{'}$ 07$^{''}$.0 & 0.051& 1.00 & 7.91 & Scaife et al. (2015) \\
RXCJ\,0431.4--6126  & A\,3266     & 04$^{\rm h}$ 31$^{\rm m}$ 24$^{\rm s}$.1  & --61$^{\circ}$ 26$^{'}$ 38$^{''}$.0 & 0.059 & 1.15 & 6.71 & This paper \\
RXCJ\,0909.1--0939  & A\,754      & 09$^{\rm h}$ 09$^{\rm m}$ 08$^{\rm s}$.4  & --09$^{\circ}$ 39$^{'}$ 58$^{''}$.0 & 0.054 & 1.06 & 6.68 & Macario et al. (2011) \\
RCXJ\,1407.8--5100  &             & 14$^{\rm h}$ 07$^{\rm m}$ 52$^{\rm s}$.5  & --51$^{\circ}$ 00$^{'}$ 33$^{''}$.0 & 0.097 & 1.81 & 6.52 & This paper \\
RXCJ\,1631.6--7507  & A\,3628     & 16$^{\rm h}$ 31$^{\rm m}$ 24$^{\rm s}$.0  & --75$^{\circ}$ 07$^{'}$ 01$^{''}$.0 & 0.105 & 1.94 & 6.49 & This paper \\
RXCJ\,2201.9--5956  & A\,3827     & 22$^{\rm h}$ 01$^{\rm m}$ 56$^{\rm s}$.0  & --59$^{\circ}$ 56$^{'}$ 58$^{''}$.0 & 0.098 & 1.82 & 5.93 & This paper \\
RXCJ\,2012.5--5649  & A\,3667     & 20$^{\rm h}$ 12$^{\rm m}$ 30$^{\rm s}$.5  & --56$^{\circ}$ 49$^{'}$ 55$^{''}$.0 & 0.056 & 1.09 & 5.77 & R\"ottgering et al. (1997)\\
RXCJ\,1358.9--4750  &             & 13$^{\rm h}$ 59$^{\rm m}$ 01$^{\rm s}$.6  & --47$^{\circ}$ 50$^{'}$ 49$^{''}$.0 & 0.074 & 1.42 & 5.44 & This paper \\
RXCJ\,1259.3--0411  & A\,1651     & 12$^{\rm h}$ 59$^{\rm m}$ 21$^{\rm s}$.5  & --04$^{\circ}$ 11$^{'}$ 41$^{''}$.0 & 0.085 & 1.61 & 5.20 & -- \\
RXCJ\,041.8--0918  & A\,85        & 22$^{\rm h}$ 01$^{\rm m}$ 56$^{\rm s}$.0  & --59$^{\circ}$ 56$^{'}$ 58$^{''}$.0 & 0.056 & 1.09 & 4.90 & Slee et al. (2001)\\
RXCJ\,0817.4-0730  & A\,644       & 08$^{\rm h}$ 17$^{\rm m}$ 24$^{\rm s}$.5  & --07$^{\circ}$ 30$^{'}$ 46$^{''}$.0 & 0.070 & 1.35 & 4.70 & This paper \\
RXCJ\,2210.3--1210 & A\,2420      & 22$^{\rm h}$ 10$^{\rm m}$ 19$^{\rm s}$.7  & --12$^{\circ}$ 10$^{'}$ 34$^{''}$.6 & 0.085 & 1.61 & 4.48 & -- \\
RXCJ\,2249.9--6425 & A\,3921      & 22$^{\rm h}$ 49$^{\rm m}$ 57$^{\rm s}$.0  & --64$^{\circ}$ 25$^{'}$ 46$^{''}$.0 & 0.094 & 1.76 & 4.34 & This paper \\
RXCJ\,2246.3--5243 & A\,3911      & 22$^{\rm h}$ 45$^{\rm m}$ 28$^{\rm s}$.7  & --53$^{\circ}$ 02$^{'}$ 08$^{''}$.0 & 0.097 & 1.81 & 4.31 & This paper \\
RXCJ\,0552.8--2103 & A\,550       & 05$^{\rm h}$ 52$^{\rm m}$ 52$^{\rm s}$.4  & --21$^{\circ}$ 03$^{'}$ 25$^{''}$.0 & 0.099 & 1.84 & 4.23 & This paper \\
PSZ1G\,018.75+23.57 &             & 17$^{\rm h}$ 02$^{\rm m}$ 22$^{\rm s}$.1  & --01$^{\circ}$ 00$^{'}$ 16$^{''}$.0 & 0.089 & 1.67 & 4.21 & This paper \\
RXCJ\,0342.8--5338 & A\,3158      & 03$^{\rm h}$ 42$^{\rm m}$ 53$^{\rm s}$.9  & --53$^{\circ}$ 38$^{'}$ 07$^{''}$.0 & 0.059 & 1.15 & 4.20 & This paper \\
RXCJ\,2154.1--5751 & A\,3822      & 21$^{\rm h}$ 54$^{\rm m}$ 09$^{\rm s}$.2  & --57$^{\circ}$ 51$^{'}$ 19$^{''}$.0 & 0.076 & 1.45 & 4.18 & This paper \\
RXCJ\,2152.4--1933 & A\,2384      & 21$^{\rm h}$ 52$^{\rm m}$ 14$^{\rm s}$.2  & --19$^{\circ}$ 42$^{'}$ 20$^{''}$.0 & 0.094 & 1.76 & 4.11 & -- \\
RXCJ\,2034.7--3548 & A\,3695      & 20$^{\rm h}$ 34$^{\rm m}$ 47$^{\rm s}$.9  & --35$^{\circ}$ 49$^{'}$ 27$^{''}$.0 & 0.089 & 1.67 & 4.06 & This paper \\
RXCJ\,1258.6--0145 & A\,1650      & 12$^{\rm h}$ 58$^{\rm m}$ 41$^{\rm s}$.1  & --01$^{\circ}$ 45$^{'}$ 25$^{''}$.0 & 0.085 & 1.61 & 4.00 & This paper \\
\hline\noalign{\smallskip}
\end{tabular}
\end{center}
\label{tab:logs}
\end{table*}

%

The data reduction was carried out using the CASA\footnote{http://casa.nrao.edu}
software. 
The data were initially flagged to excise RFI and the bandpass, i.e. antenna 
complex gains as a function of frequency, was derived for PKS\,1934--68 and 
then applied to the target visibilities. Calibrated visibilities were Fourier 
transformed to an image using the w-projection algorithm 
(Perley \& Cornwell 1991) to account for the array non coplanarity. 
Briggs weights 
(Briggs 1995) with robust parameter 0.5 were used in imaging in order to 
achieve a tradeoff between angular resolution and sensitivity to diffuse 
emission. 
Images were deconvolved using the Cotton--Schwab algorithm until the first 
negative component was found. Such sky model was then used for a phase 
selfcalibration where antenna phase variations were solved for in a 5~minute 
interval. After selfcalibration, gain solutions were applied and visibility data 
Fourier transformed into an image that was deconvolved down to a 1~mJy 
threshold to obtain the final images. For selfcalibration and deconvolution 
purposes, a $4^\circ \times 4^\circ$ field of view was 
imaged, although we show here only the inner $2^\circ \times 2^\circ$ 
deconvolved images in Appendix A. 
We conservatively estimate that residual amplitude calibration errors 
are smaller than 5\%.

Given the limited KAT--7 angular resolution, we expect the noise in the central 
portion of the images to be a combination of the thermal and confusion 
contributions respectively (calibration errors play a minor role as the 
dynamic range of the images is of the order of a few thousands to one at most).
We estimated noise values (see Table~\ref{tab:obs}) as the rms in areas that appear void of detectable sources, finding them at the $0.3 - 1$~mJy
level across the whole target sample and steadily decreasing to 
$130 - 250$~$\mu$Jy at the edge of the $4^\circ \times 4^\circ$ images, indeed 
indicating that confusion contributes to the noise budget at the image centre. 
An estimate of the confusion level can be obtained as (Condon 1987):
\begin{equation}
\sigma_c \approx 0.2 \, {\rm mJy~beam}^{-1} \, {\left ( \frac{\nu}{\rm GHz} \right )}^{-0.7} \left ( \frac{\theta}{{\rm arcmin}} \right )^2,
\end{equation}
that gives a confusion noise at the $\sim$~0.7~mJy level in the KAT--7 images. 
A similar value could be estimated from the independent measurements by 
Garrett et al. (2000), indicating that our observations are likely approaching 
the confusion level at the image centre.


\begin{table*}
\caption[]{Observational details, RH powers and upper limits (UL). Note: The Triangulum Australis RH measurement is derived from observations at 1.33~GHz.} 
\begin{center}
\begin{tabular}{lcccc}
\hline
\hline\noalign{\smallskip}
Cluster name & Obs. time & Noise rms         & Angular    & RH powers and UL \\
             & (hour)    & (mJy~beam$^{-1}$) & resolution & ($\log{P_{\rm 1.4}}$, Watt~Hz$^{-1}$) \\
\hline\noalign{\smallskip}
Triangulum           &  12 & 0.73 & 2.9$^{\prime}\times2.7^{\prime}$ & 23.73 \\
A\,3266              &  7  & 0.72 & 2.7$^{\prime}\times2.3^{\prime}$ & -\\
RCXJ\,1407.8--5100   &  6  & 0.47 & 2.9$^{\prime}\times2.2^{\prime}$ & - \\
A\,3628	             &  9  & 1.0  & 2.8$^{\prime}\times2.4^{\prime}$ & - \\
A\,3827              &  6  & 0.53 & 2.8$^{\prime}\times2.3^{\prime}$ & - \\
RXCJ\,1358.9--4750   &  6  & 0.50 & 2.5$^{\prime}\times2.2^{\prime}$ & $<$24.00\\
A\,644               &  5  & 0.66 & 2.6$^{\prime}\times2.4^{\prime}$ & $<$23.95\\
A\,3921              &  5  & 0.42 & 2.8$^{\prime}\times2.1^{\prime}$ & $<$24.22\\
A\,3911              &  7  & 0.51 & 2.5$^{\prime}\times2.4^{\prime}$ & - \\
A\,550               &  6  & 0.77 & 2.5$^{\prime}\times2.3^{\prime}$ & $<$24.27\\
PSZ1G\,018.75+23.57  &  6  & 0.39 & 2.8$^{\prime}\times2.4^{\prime}$ & 24.15 \\
A\,3158              &  8  & 0.30 & 2.4$^{\prime}\times2.2^{\prime}$ & $<$23.79\\
A\,3822              &  5  & 0.48 & 2.8$^{\prime}\times2.0^{\prime}$ & $<$24.02 \\
A\,3695              &  7  & 0.90 & 2.6$^{\prime}\times2.4^{\prime}$ & - \\
A\,1650              &  5  & 0.52 & 2.7$^{\prime}\times2.4^{\prime}$ & $<$24.12 \\
\hline\noalign{\smallskip}
\end{tabular}
\end{center}
\label{tab:obs}
\end{table*}


\begin{figure*}
\centering
\includegraphics[width=1\columnwidth]{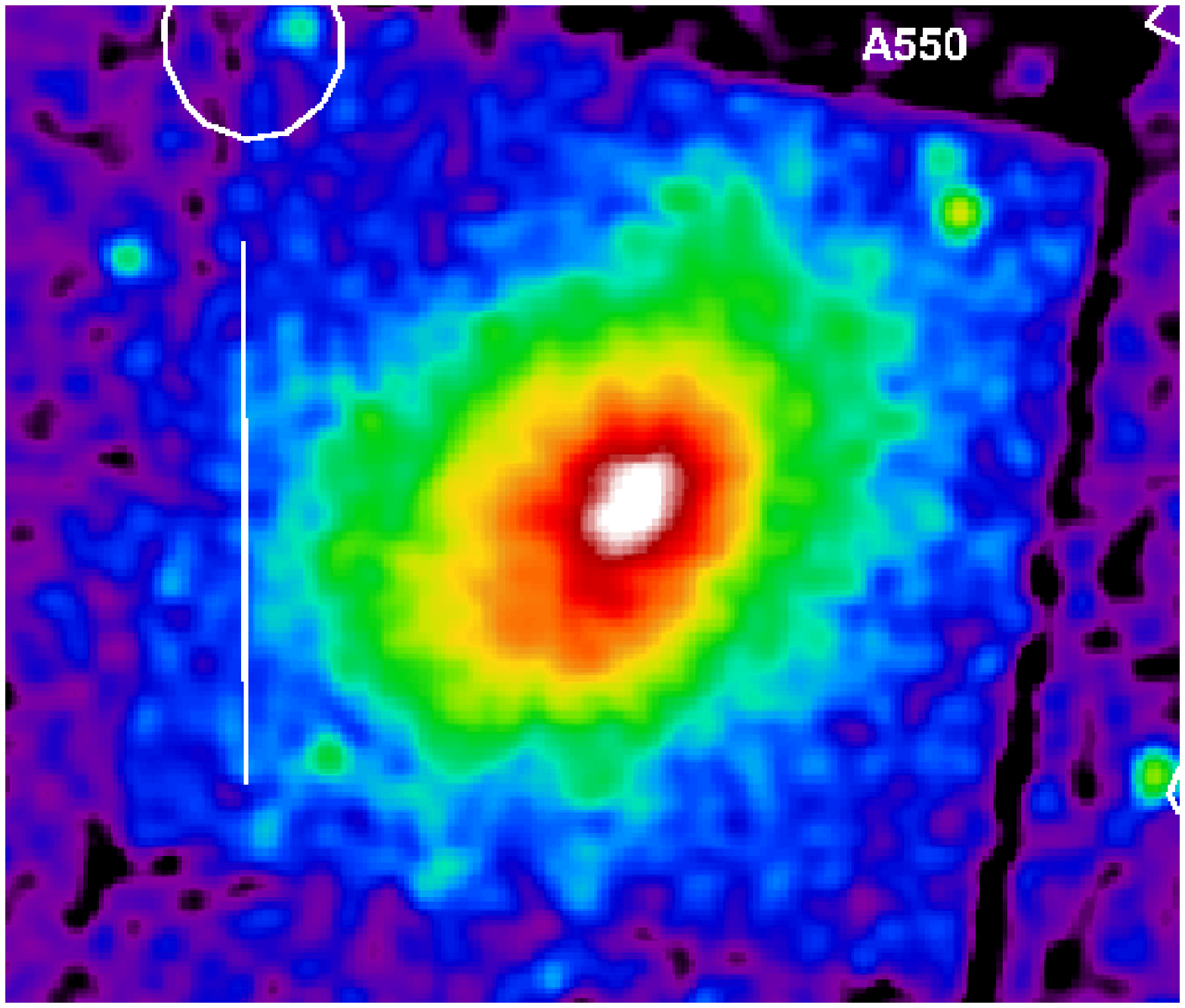}
\includegraphics[width=1\columnwidth]{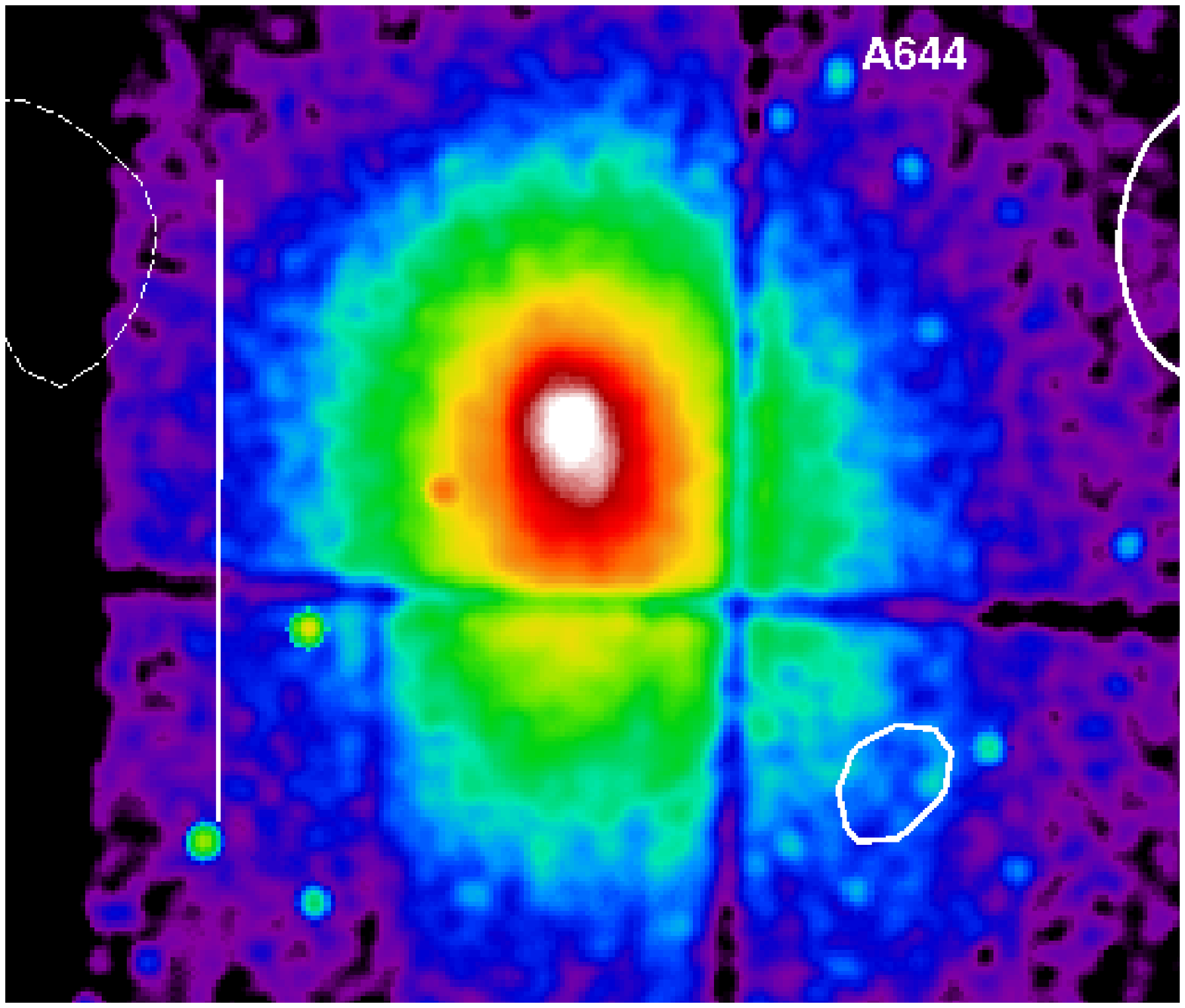}
\includegraphics[width=1\columnwidth]{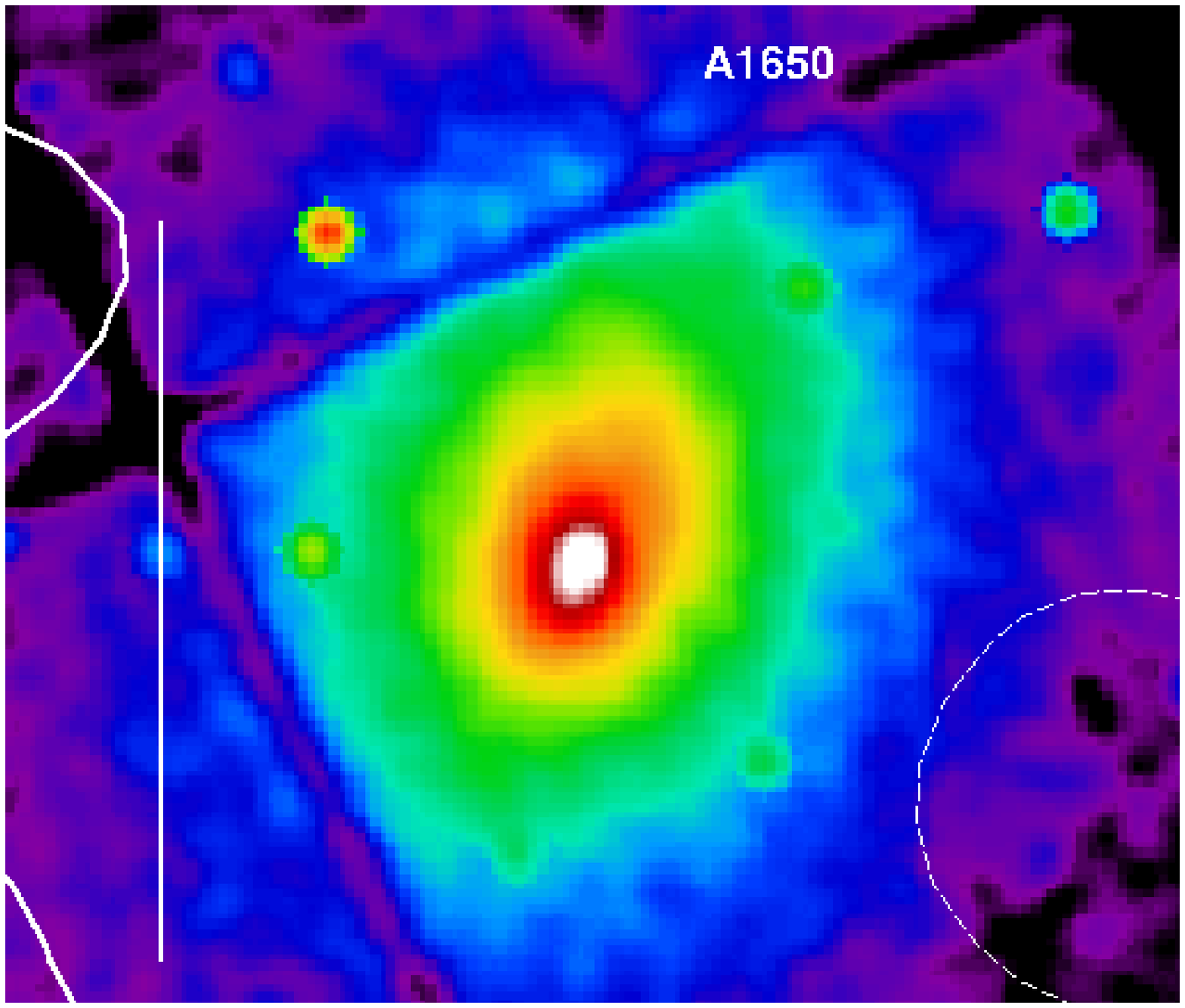}
\includegraphics[width=1\columnwidth]{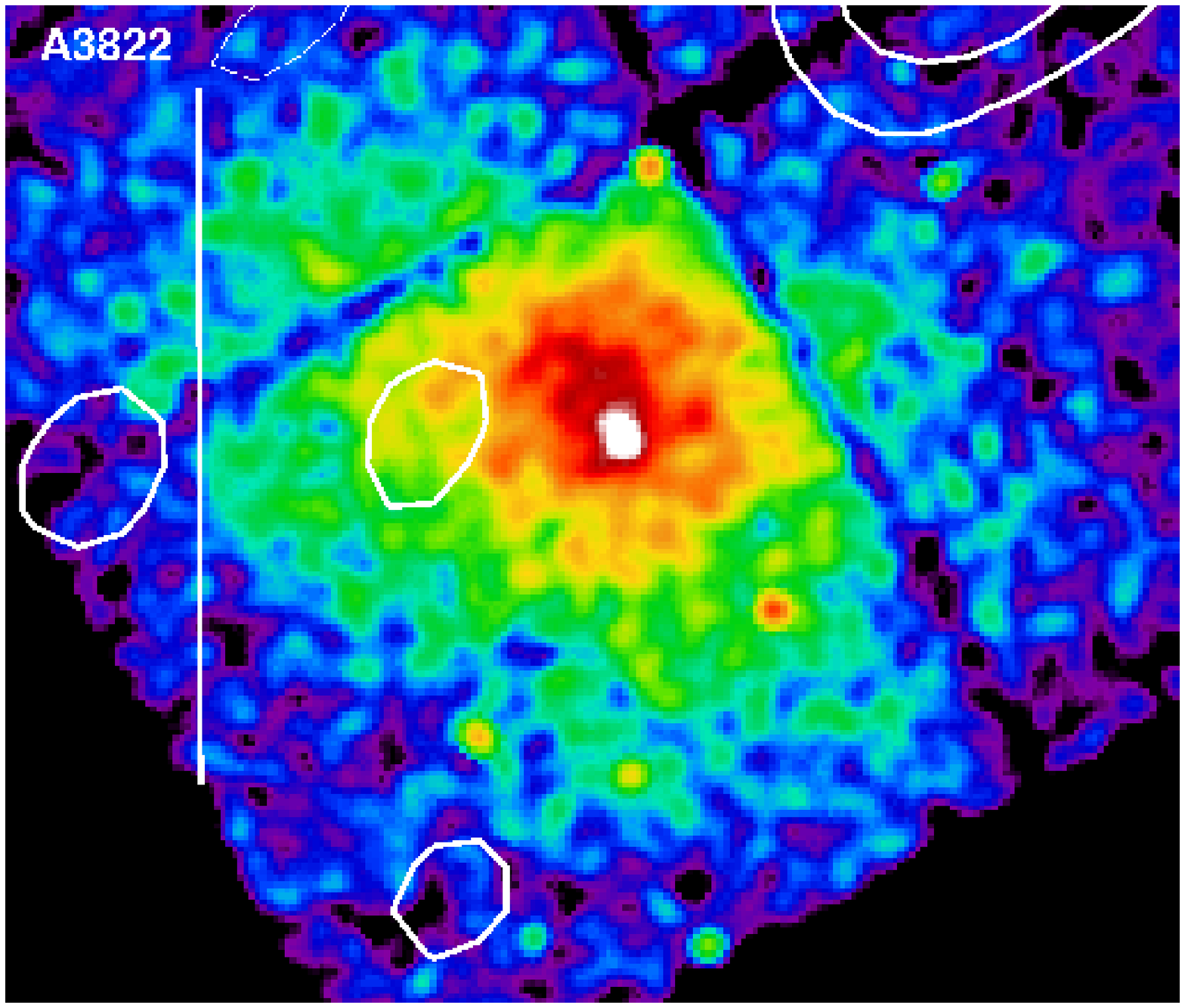}
\includegraphics[width=1\columnwidth]{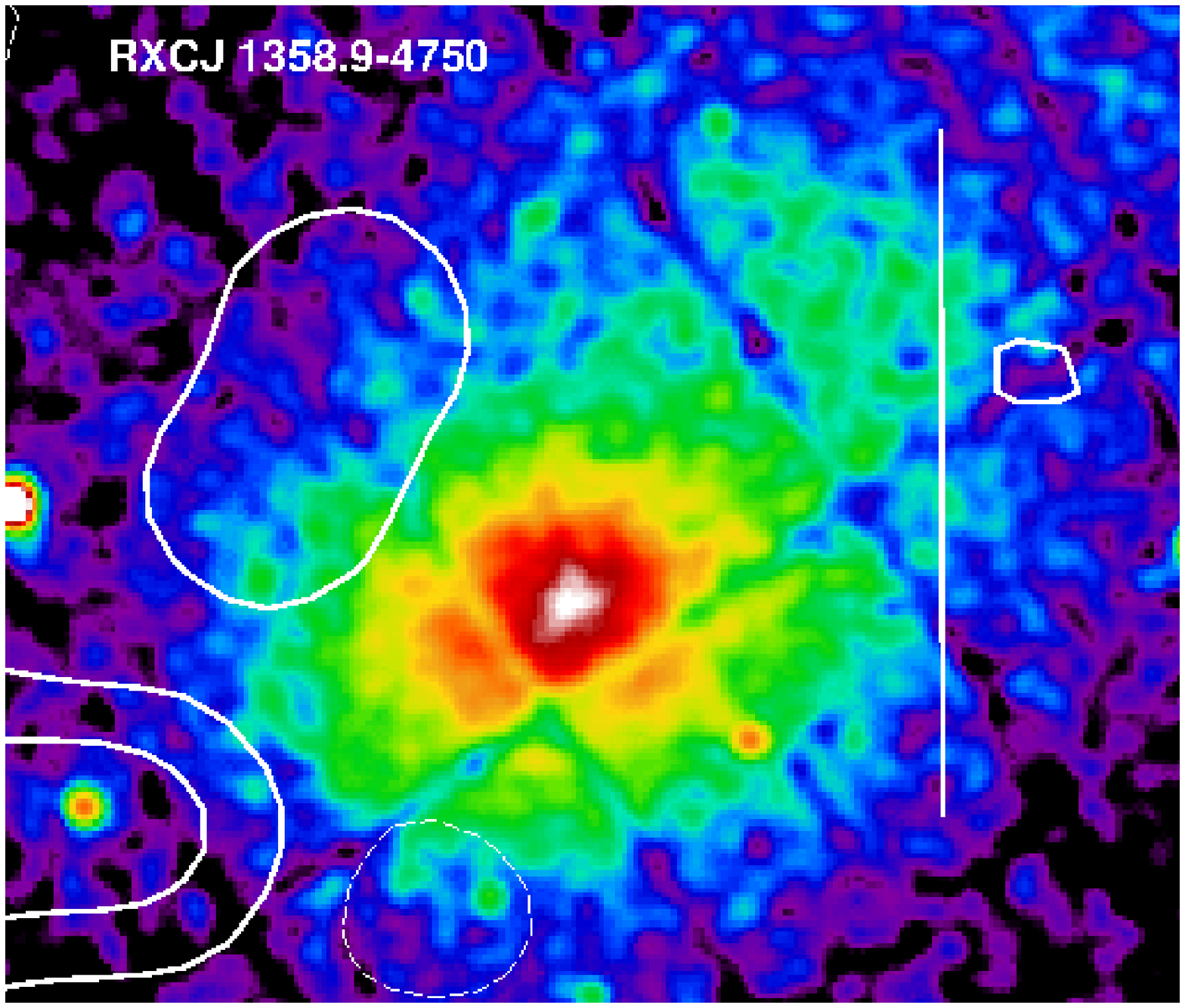}
\caption{1.86~GHz KAT--7 radio contours of clusters that show no central radio emission, overlaid on the X--ray images. Top panels: A\,550 (left) on {\it XMM--Newton} and A\,644 (right) on {\it Chandra}. Middle panels: A\,1650 (left) and A\,3822 (right) on {\it Chandra}. Bottom panel: RXCJ1358--4750 on {\it XMM--Newton}. Radio contours are drawn at -2.5, 2.5, 10 and 40~mJy~beam$^{-1}$ (A\,550, A\,644 and A\,1650) and -1.5, 1.5, 6, 24~mJy~beam$^{-1}$ (A\,3822 and RXCJ\,1358.9--4750). Positive and negative contours are drawn with solid and dashed lines respectively. Throughout the paper, KAT--7 radio contours are drawn from images not corrected by the primary beam, although all the reported flux densities are. The vertical white bar indicates a 800~kpc size.}
\label{fig:noradio}
\end{figure*}
%
\begin{figure*}
\centering
\includegraphics[width=1\columnwidth]{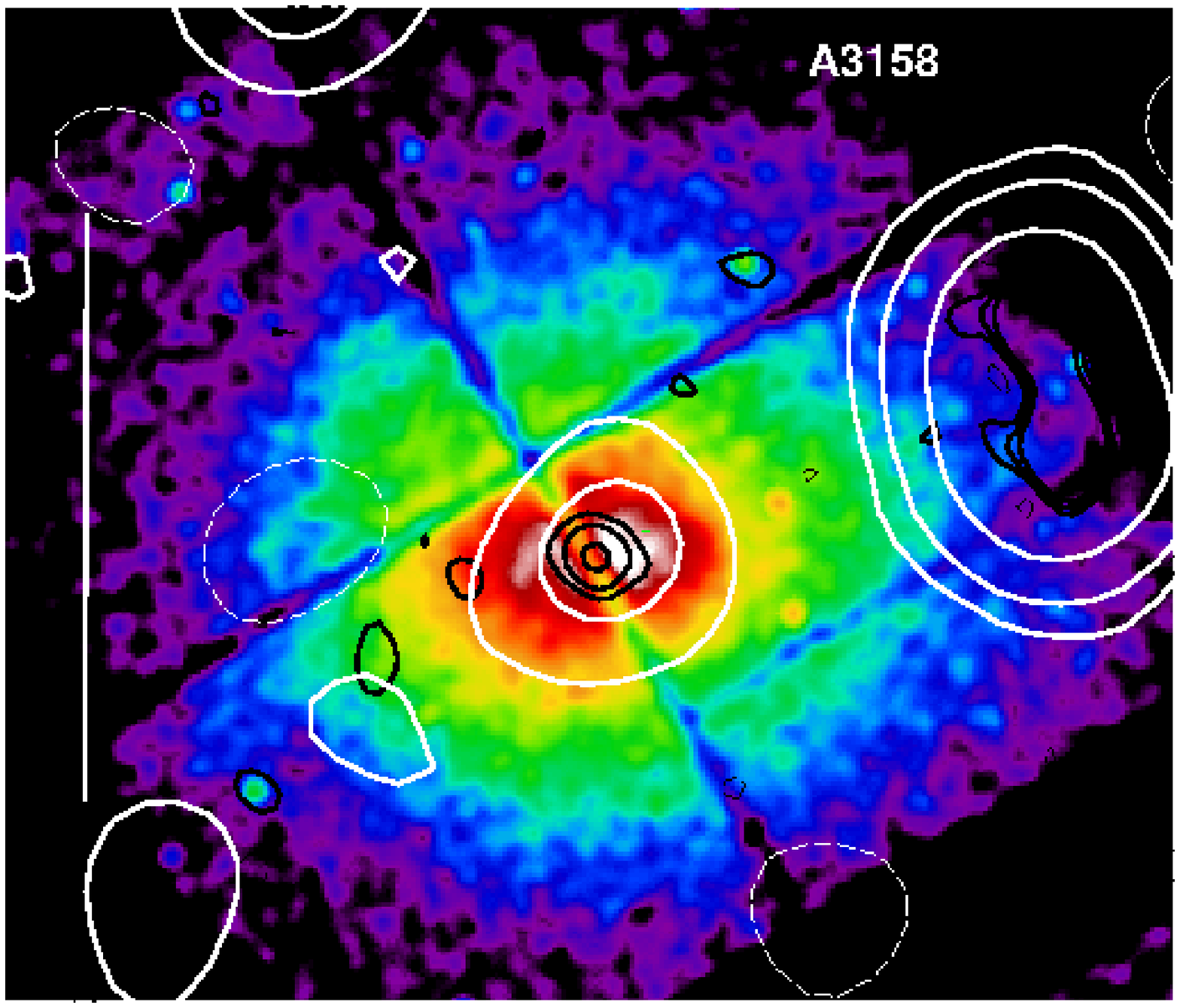}
\includegraphics[width=1\columnwidth]{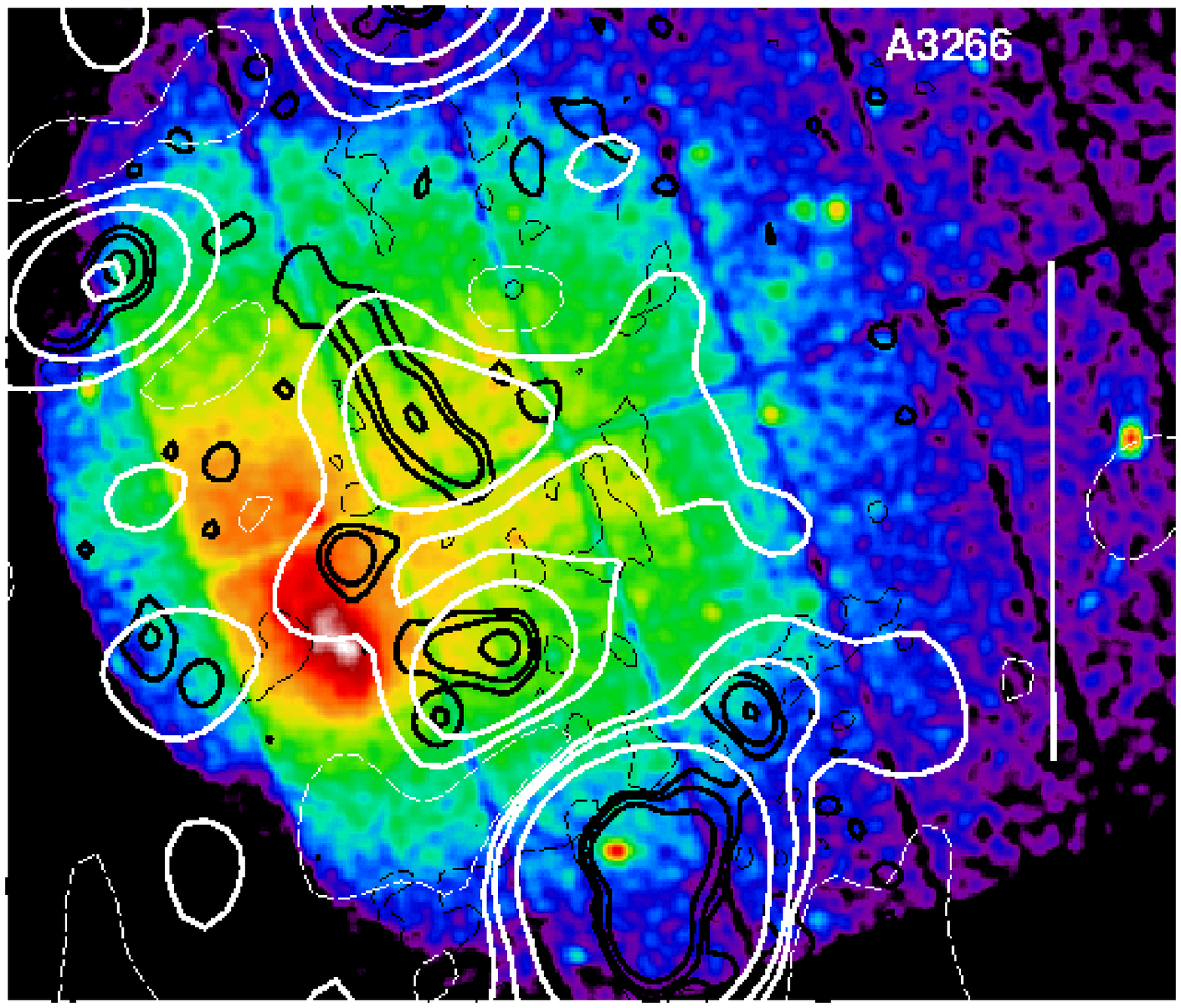}
\includegraphics[width=1\columnwidth]{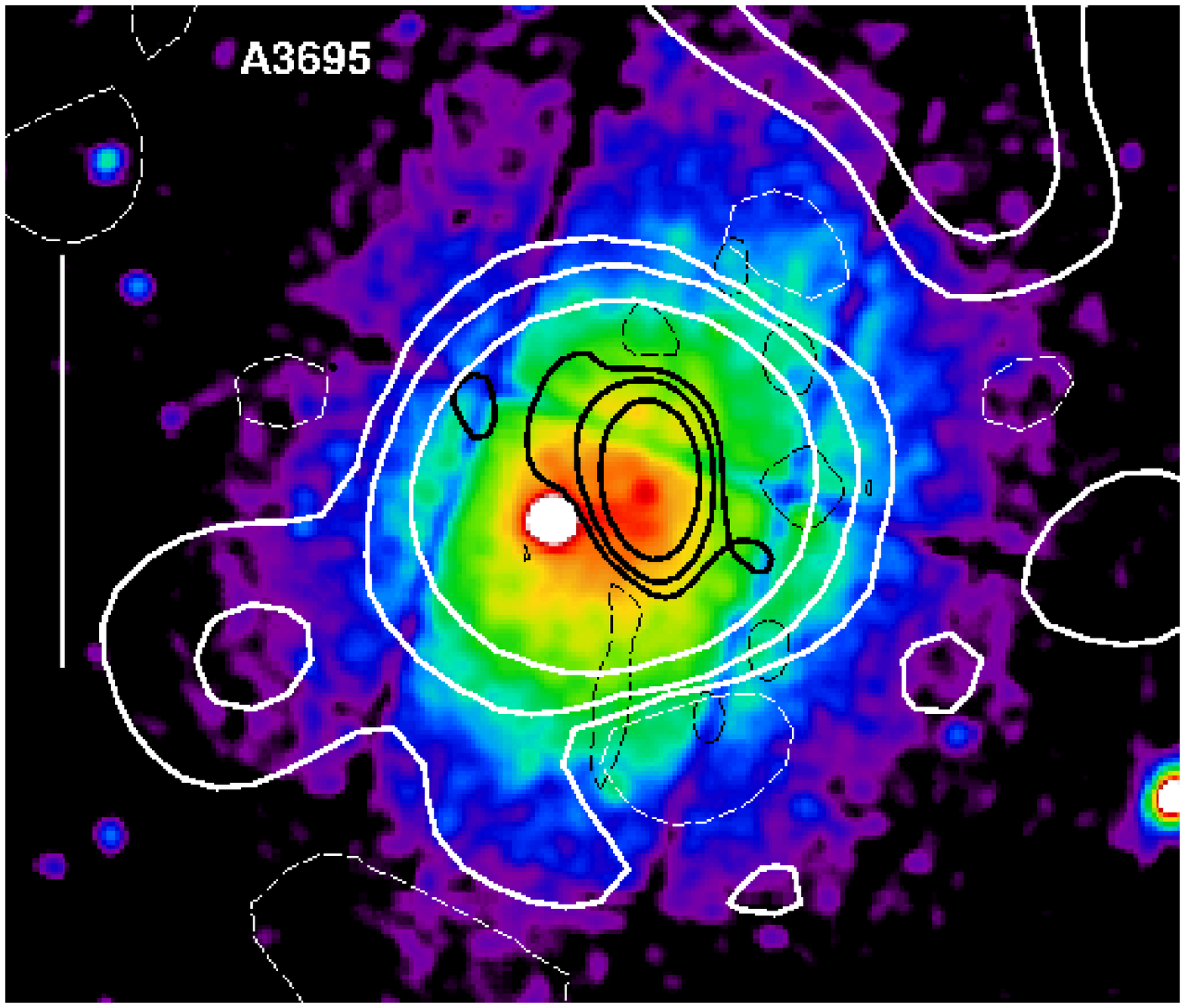}
\includegraphics[width=1\columnwidth]{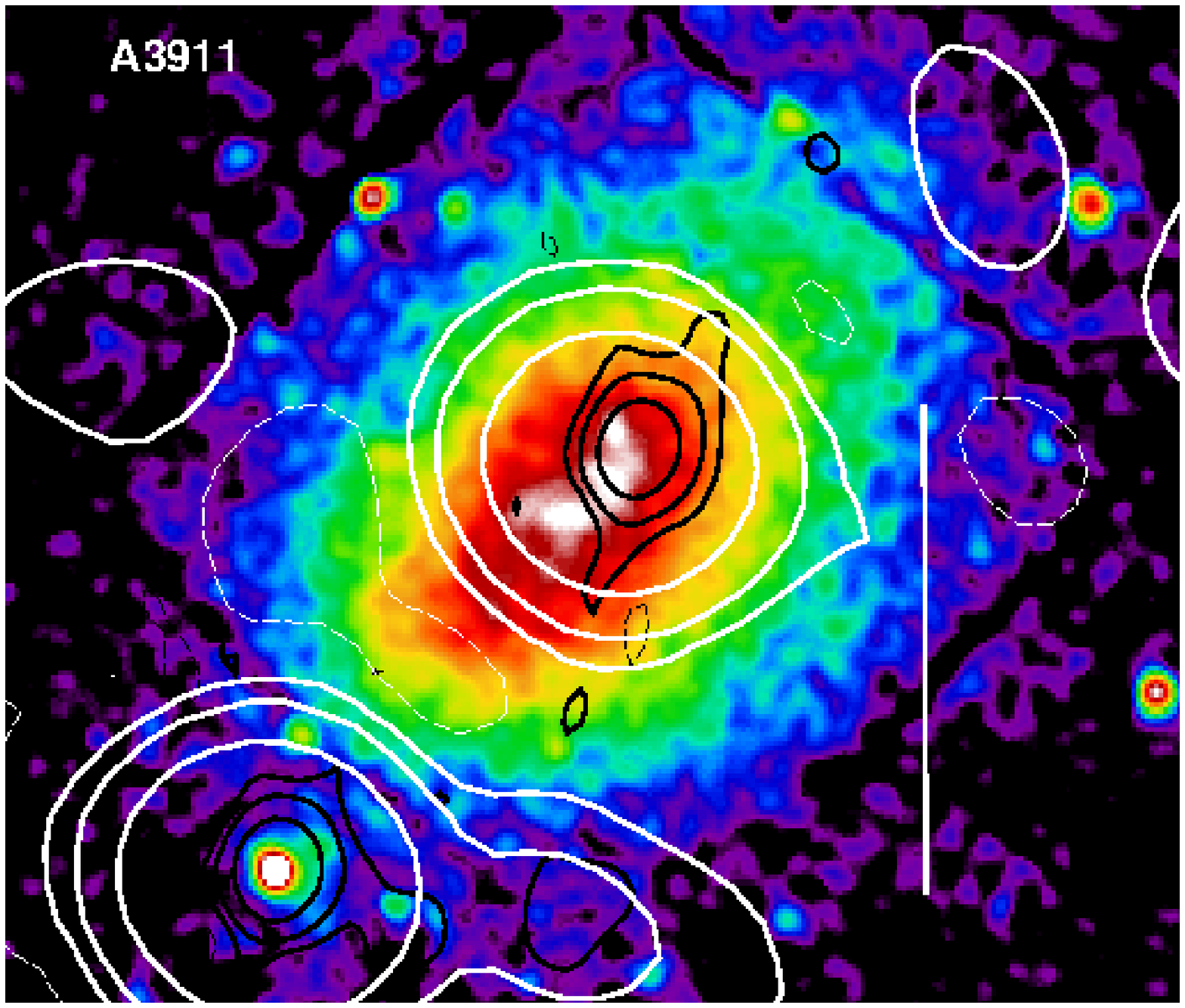}
\caption{1.86~GHz KAT--7 radio contours (white) of clusters whose central radio emission is a blend of discrete sources, overlaid on the X--ray images and on the 843 MHz contours from SUMSS (black). The vertical white bar indicates a 800~kpc size. Top panels: A\,3158 (left) on {\it Chandra}, KAT--7 contours are drawn at -1, 1, 4, 16~mJy~beam$^{-1}$, SUMSS contours are drawn at $\pm$~3, 6, 12~mJy~beam$^{-1}$;
A\,3266 (right) on {\it XMM--Newton}, KAT--7 contours are drawn at $\pm$~2.5, 10, 40~mJy~beam$^{-1}$, SUMSS contours are drawn at $\pm$~3, 12, 48~mJy~beam$^{-1}$. Bottom panels: A\,3695 (left) on {\it XMM--Newton}, KAT--7 contours are drawn at $\pm$~3, 12, 48~mJy~beam$^{-1}$, SUMSS contours are drawn at $\pm$~20, 80, 320 ~mJy~beam$^{-1}$; A\,3911 (right) on {\it Chandra}, KAT--7 contours are drawn at $\pm$~1.5, 6, 24~mJy~beam$^{-1}$, SUMSS contours are drawn at $\pm$~6, 24, 96~mJy~beam$^{-1}$.}
\end{figure*}
\begin{figure*}
\ContinuedFloat
\captionsetup{list=off,format=cont}
\centering
\includegraphics[width=1\columnwidth]{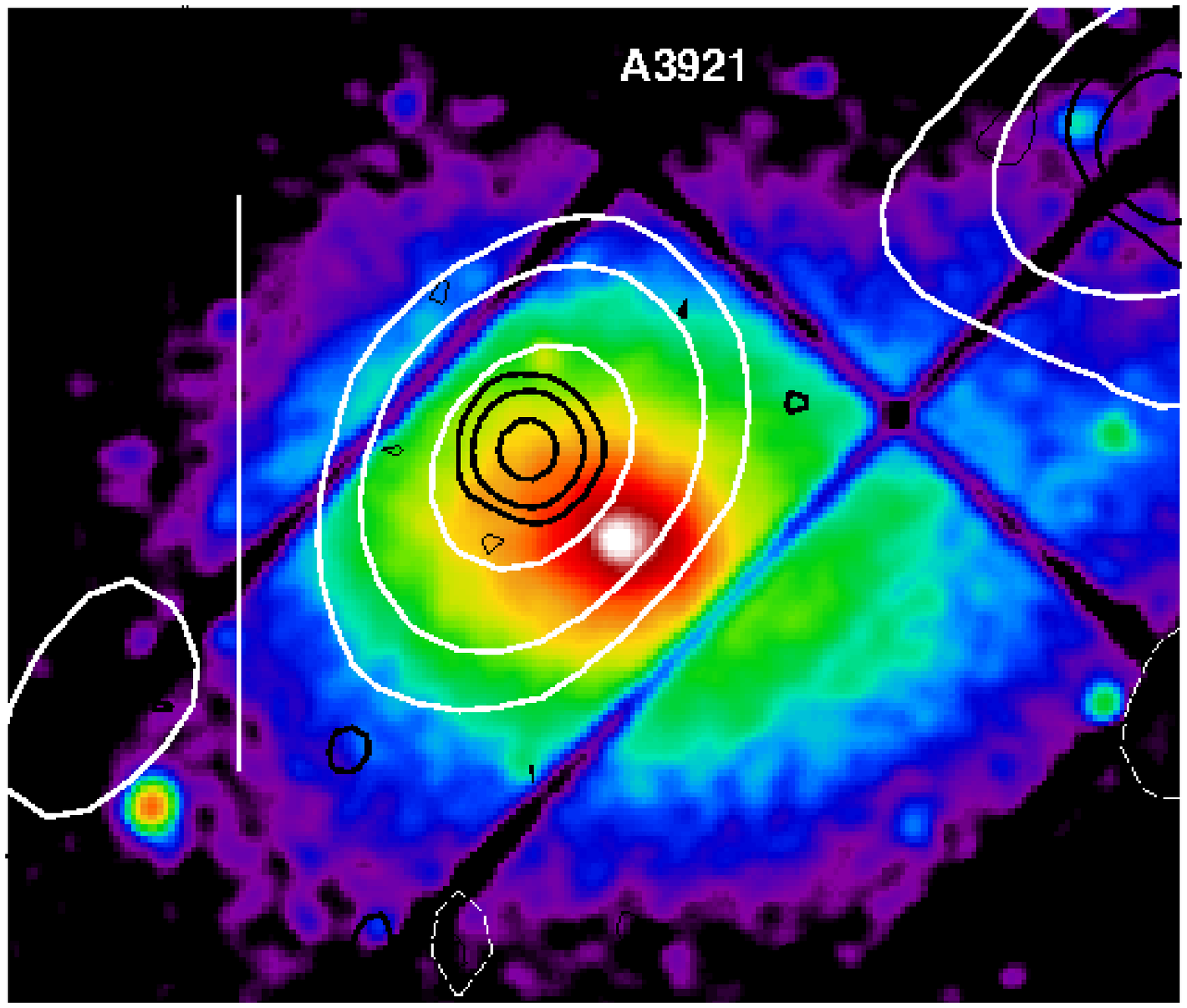}
\includegraphics[width=1\columnwidth]{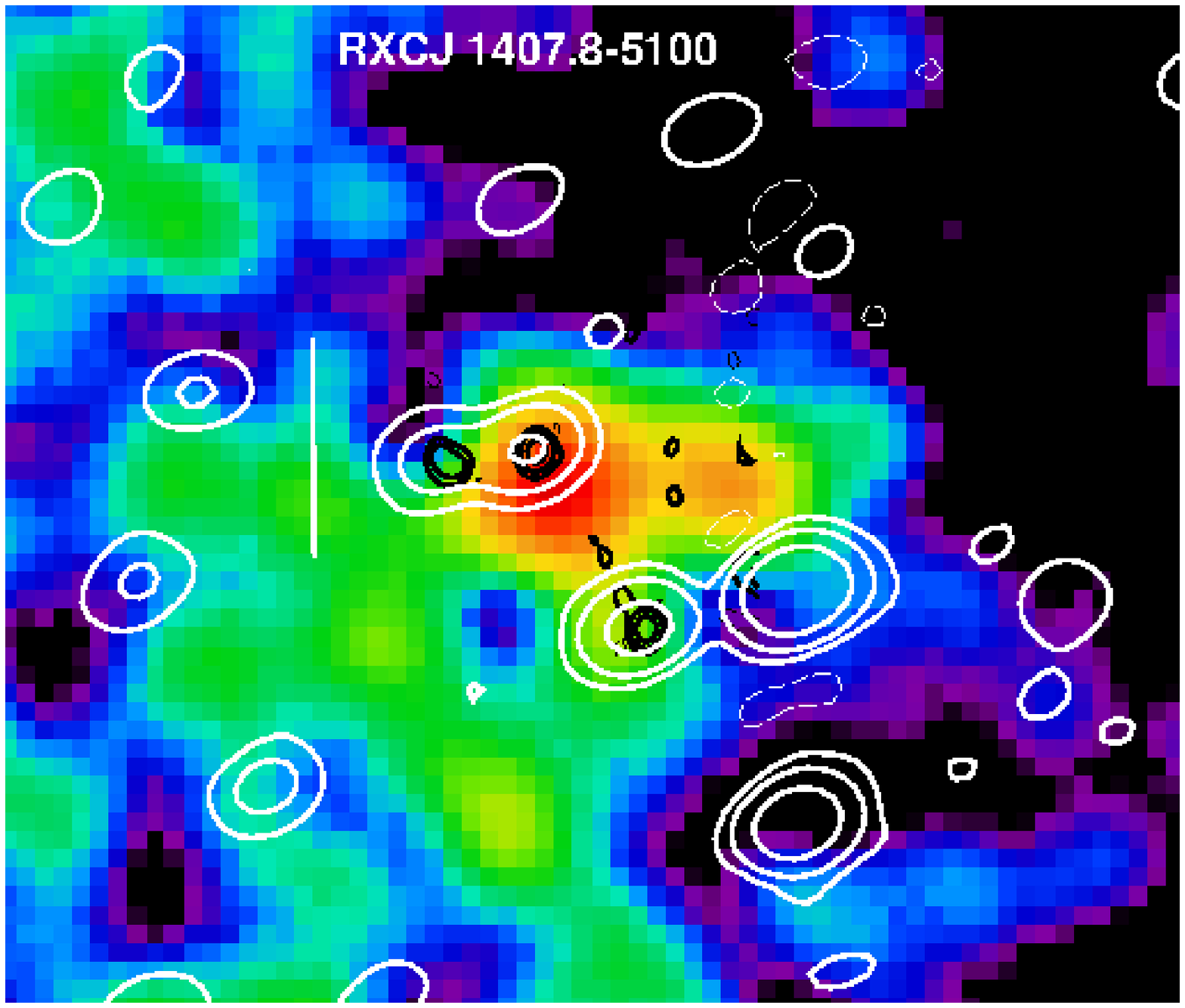}
\includegraphics[width=1\columnwidth]{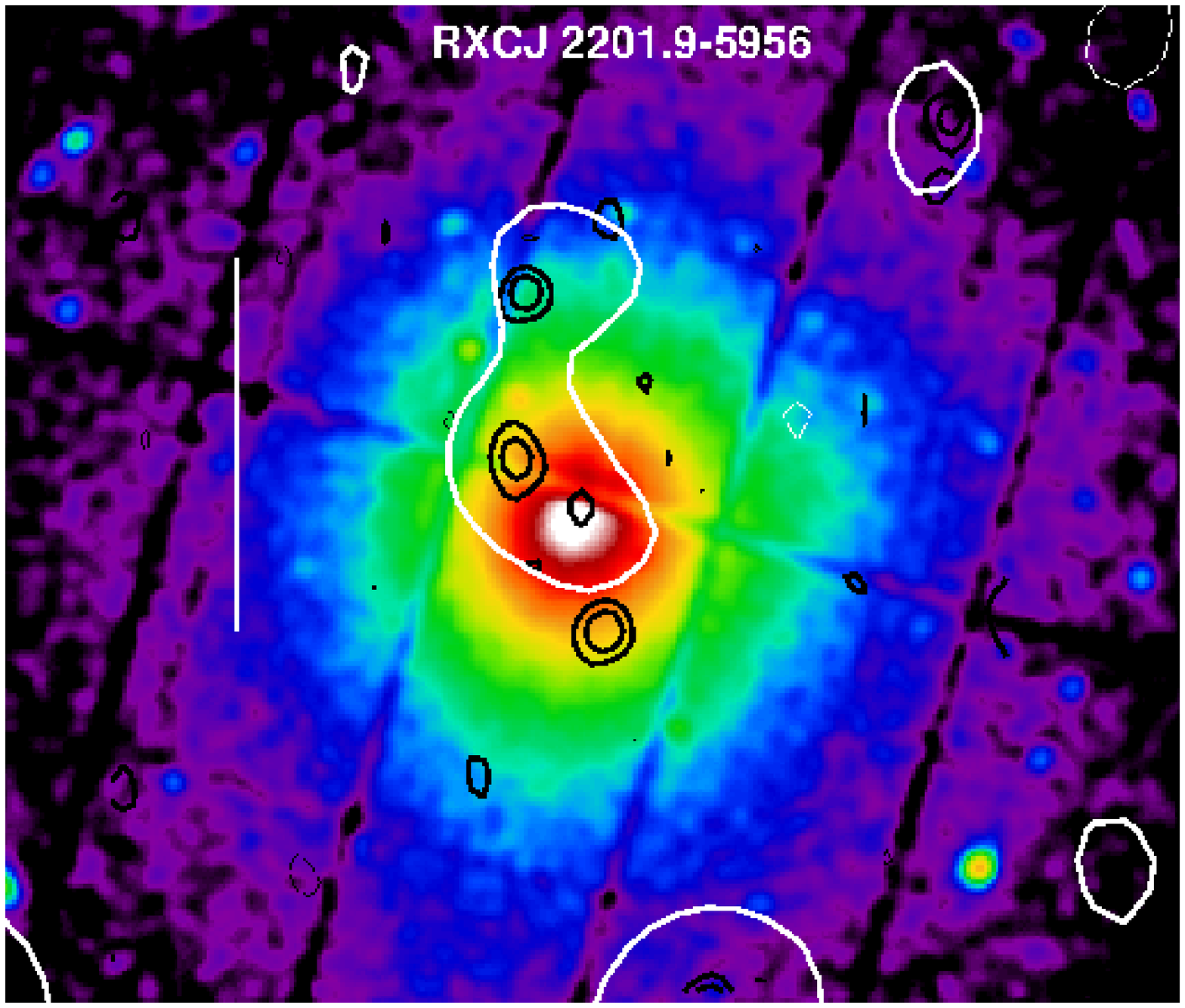}
\caption{Continued. Top panels: A\,3921 (left) on {\it Chandra}, KAT--7 contours are drawn at $\pm$~1.5, 6, 24~mJy~beam$^{-1}$, SUMSS contours are drawn at $\pm$~3, 12, 48~mJy~beam$^{-1}$;  RXCJ\,1407.8--5100 (right) on ROSAT, KAT--7 contours are drawn at $\pm$~3, 12, 48~mJy~beam$^{-1}$, SUMSS contours are drawn at $\pm$~4, 16, 64~mJy~beam$^{-1}$. Bottom panel: RXCJ\,2201.9--5956 on {\it XMM--Newton}, KAT--7 contours are drawn at $\pm$~2~mJy~beam$^{-1}$, SUMSS contours are $\pm$~2.5, 5~mJy~beam$^{-1}$.}
\label{fig:strong}
\end{figure*}
%
\begin{figure*}
\centering
\includegraphics[width=1\columnwidth]{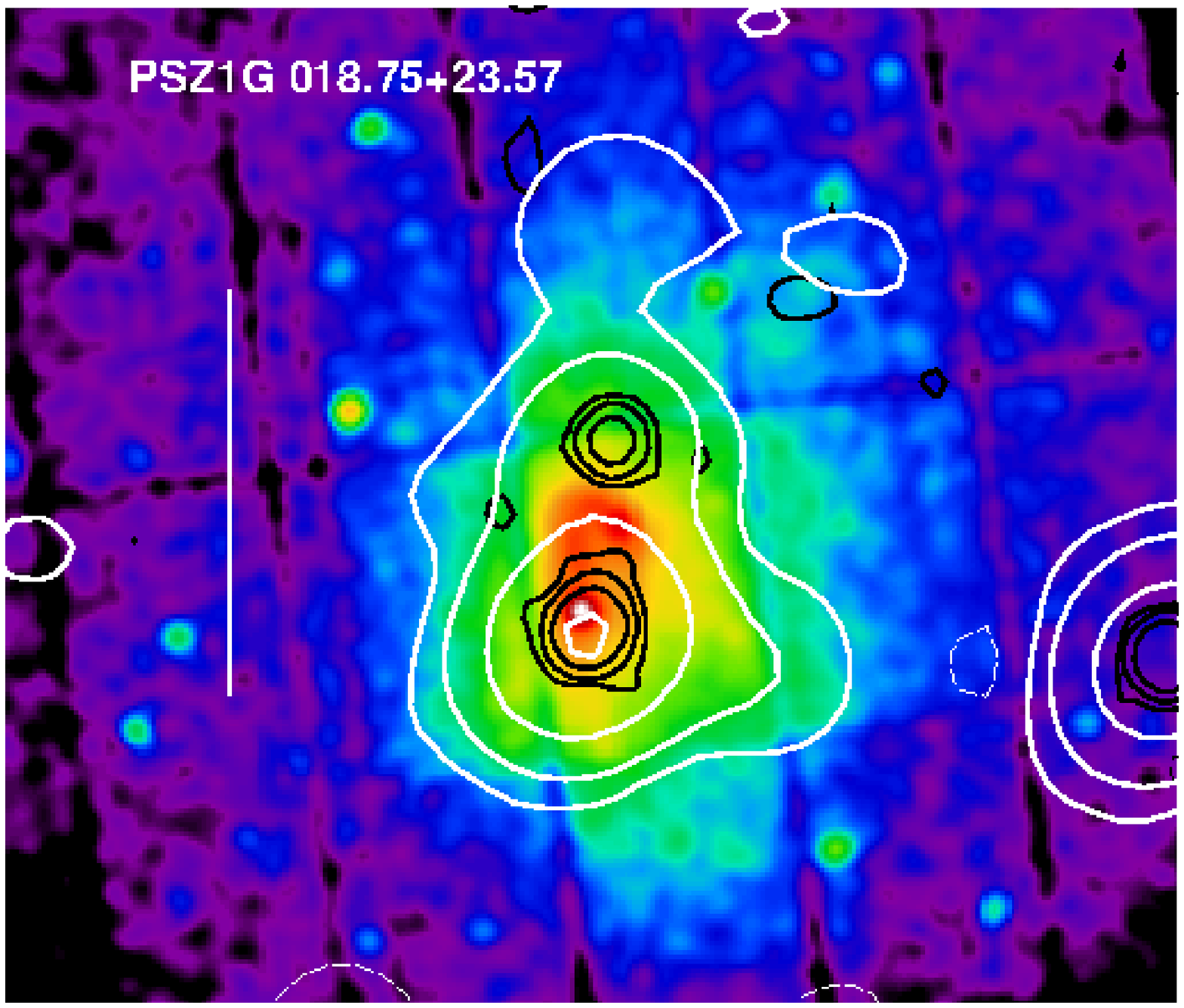}
\includegraphics[width=1\columnwidth]{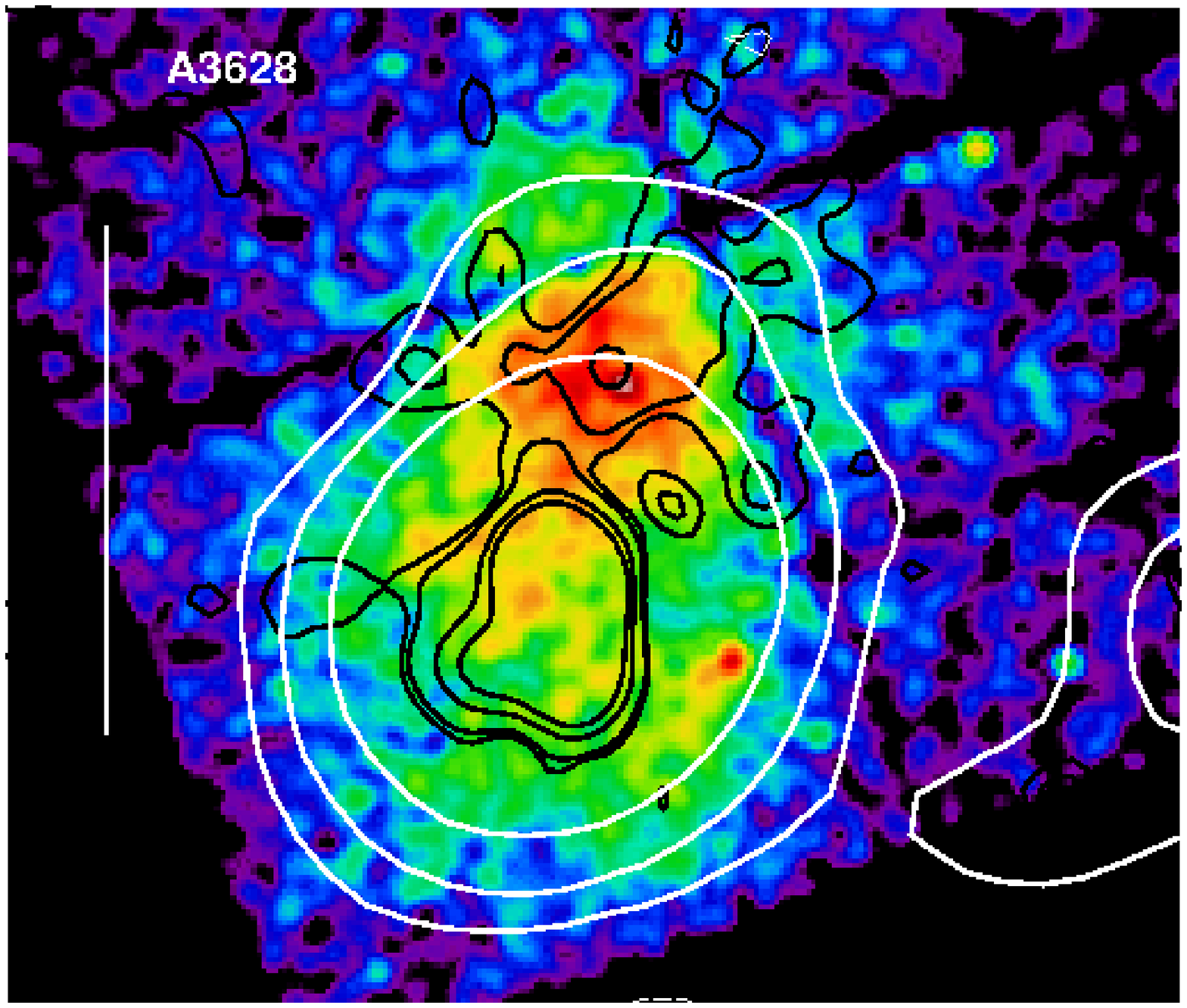}
\includegraphics[width=1\columnwidth]{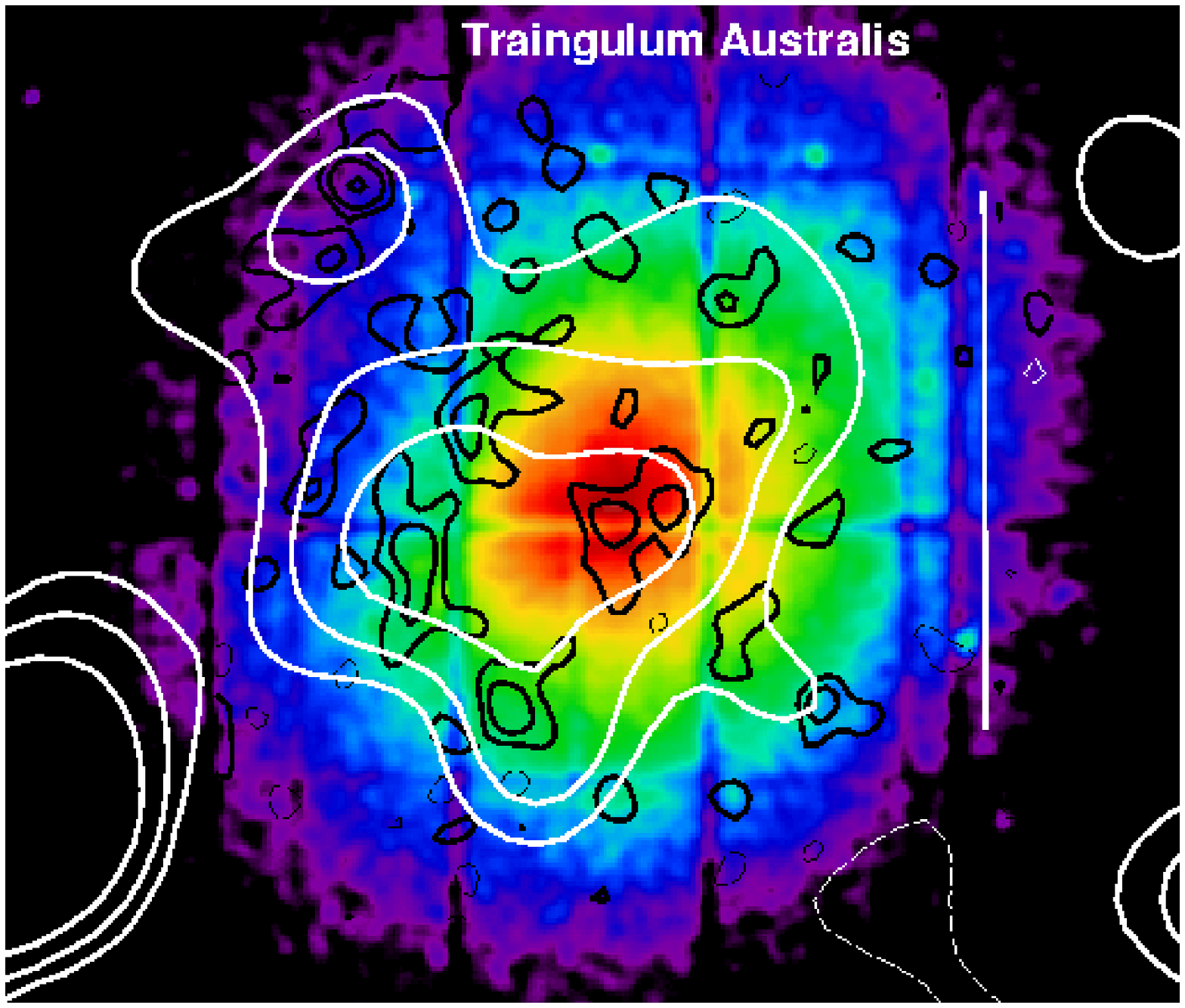}
\caption{1.86~GHz KAT--7 radio contours (white) of clusters with candidate diffuse radio halo emission, overlaid on the X--ray images. The vertical white bar indicates a 800~kpc size. From top to bottom: PSZ1G\,018.75+23.57 on {\it XMM--Newton}, KAT--7 contours are drawn at $\pm$~2.5, 10, 40, 160~mJy~beam$^{-1}$, NVSS contours (black) are drawn at $\pm$~1.5, 6, 24~mJy~beam$^{-1}$;
A\,3628 on {\it Chandra}, KAT--7 contours are drawn at $\pm$~2, 8, 32~mJy~beam$^{-1}$, SUMSS contours (black) are drawn at $\pm$~4, 8, 32, 64~mJy~beam$^{-1}$; Triangulum Australis on {\it XMM--Newton}, KAT--7 contours are drawn $\pm$~2.5, 5, 10~mJy~beam$^{-1}$, SUMSS contours (black) are drawn at $\pm$~2, 4, 8~mJy~beam$^{-1}$.}
\label{fig:candidate}
\end{figure*}
%
\section{Images and upper limits}
\label{sec:sources}

Figures~\ref{fig:noradio}, \ref{fig:strong} and \ref{fig:candidate} display
the KAT--7 radio emission from the central region of the 14 clusters in 
Table~\ref{tab:obs}, overlaid on the best available smoothed X--ray images
(see Figure captions). For all clusters the X--ray images presented have
been obtained from the public archives, with no further analysis.

The KAT--7 angular resolution is of the order of 2.5~arcmin (see Table~\ref{tab:obs}),
which makes it difficult to disentangle the contribution 
of discrete and diffuse sources if radio emission is detected in the central 
region of the cluster, unless high resolution images at a similar frequency 
are available in the literature.

In the following sections we describe results for each individual target.

\subsection{Clusters with no emission at their centre}
\label{sec:noradio}

In five clusters we could not detect radio emission above our sensitivity 
limits. These are A\,1650, A\,3822, A\,550, A\,644 and RXCJ\,1358.9--4750 
and their images are reported in Figure~\ref{fig:noradio}.

\subsection{Clusters with blended central radio emission}
\label{sec:scrs}

In seven clusters radio emission is detected at the centre, most likely 
blended emission from individual galaxies, as reported here below.
The 1.86~GHz KAT--7 contours for these targets are shown in white in Figure~\ref{fig:strong}, 
overlaid on the X--ray image and on the 843~MHz radio contours from SUMSS (black).

\begin{itemize}

\item {\bf A\,3158}: We detect a barely resolved faint radio source at the cluster centre, whose flux density is S$_{\rm 1.86}=7.6\pm 0.4$~mJy. 
Johnston-Hollitt et al. (2008) present a radio--optical study of 
A\,3158, where they identify all the compact radio sources brighter than 
0.15~mJy, with an angular resolution of $\sim 7$~arcsec and $\sim 4$~arcsec 
at 1.4 and 2.5~GHz respectively. They identify an unresolved source 
at RA$_{\rm J2000}$ = $3^{\rm h} 42^{\rm m} 53^{\rm s}$, 
DEC$_{\rm J2000}$ = $-53^\circ 36' 53''$ with a S$_{\rm 1.4} = 7.88$~mJy flux 
density and $\alpha = 0.8$ spectral index (S$_{\nu} \propto \nu^{-\alpha}$). Scaled to 
1.86~GHz, this source matches the flux density we detect with KAT--7
at the cluster centre. 

\item {\bf A\,3266}: Two sources are located within a few arcminutes from the X-ray 
centre of the cluster. 
Their flux density is S$_{\rm 1.86}=53.1\pm2.7$~mJy and
S$_{\rm 1.86}=40.9\pm2.0$~mJy respectively for the northern and the southern
source. The spatial correlation between the KAT--7 and SUMSS peaks suggests that most of the KAT--7 emission is associated with discrete radio sources.

\item {\bf A\,3695}: A strong point source (S$_{\rm 1.86}=1.33\pm0.07$~Jy) is located 
at the cluster centre. 
The source is point--like also in the NVSS, with a peak flux density of 
S$_{\rm 1.4} \sim 1.4$~Jy. 
The 843~MHz image shows an extension at about $\sim 40^{\circ}$ towards 
North--East, while the KAT--7 emission is marginally resolved, with a position 
angle of $\sim 120^{\circ}$. Due to the lack of high resolution information, 
the nature of the central emission in this cluster remains unclear.

\item {\bf A\,3911}: The central point--like source has a flux density  
S$_{\rm 1.86}=127.0\pm6.4$~mJy. 
A single point source is found in SUMSS, coincident with the northernmost 
part of the X--ray cluster emission. The lack of high resolution images 
either from public surveys or from the literature does not allow to properly 
classify the emission visible on the KAT--7 image.

\item {\bf A\,3921}: A single source is detected slightly offset from
the cluster X--ray emission, with flux density S$_{\rm 1.86}=52.4 \pm 2.6$~mJy.
Ferrari et al. (2006) 
present 1.3 and 2.4~GHz observations of this cluster, with 12 and 8~arcsec 
resolution respectively, where this source (SUMSS J\,225007-646239) appears unresolved.
They measure a flux density of S$_{1.34}$~=~63.8 mJy, and derive a spectral 
index $\alpha$~=~0.72 between 1.3 and 2.4~GHz. Once the flux density is 
extrapolated at 1.86~GHz using this spectral index, and the source is 
subtracted from the KAT--7 $uv$-data, the residual image is virtually noise 
limited.

\item {\bf RXCJ\,1407.8--5100}: this cluster hosts a faint double source aligned in the 
east--west direction. 
Unfortunately, only a ROSAT All Sky Survey image 
(RA$_{\rm J2000}=14^h 08^m 2.0^s$, 
DEC$_{\rm J2000}=-50^\circ 59^{\prime}3^{\prime\prime}$) 
is available for the X--ray emission of the cluster, whose peak is 
almost coincident with the radio peak. 
Comparison
with the SUMSS image shows two compact sources at the location of the
KAT--7 peaks, suggesting that the central radio emission is a blend of 
individual sources.

\item {\bf RXCJ\,2201.9--5956 (A\,3827)}: A faint and diffuse source is detected at the cluster centre at $\sim 3$$\sigma$ level, with flux density S$_{\rm 1.86}=8.8\pm0.4$~mJy. From the overlay shown in Figure~\ref{fig:strong}, it is clear that the 1.86~GHz KAT--7 emission follows the alignment of the individual sources in the SUMSS image, and it is most likely a blend.

\end{itemize}

\subsection{Candidate cluster-scale diffuse emission}
\label{sec:candidate_radio_halos}

Three galaxy clusters show hints of extended emission at their centres, which 
makes them potential candidates to host giant RHs 
(see Figure~\ref{fig:candidate}).

\begin{figure}
\centering
\includegraphics[width=1.0\columnwidth]{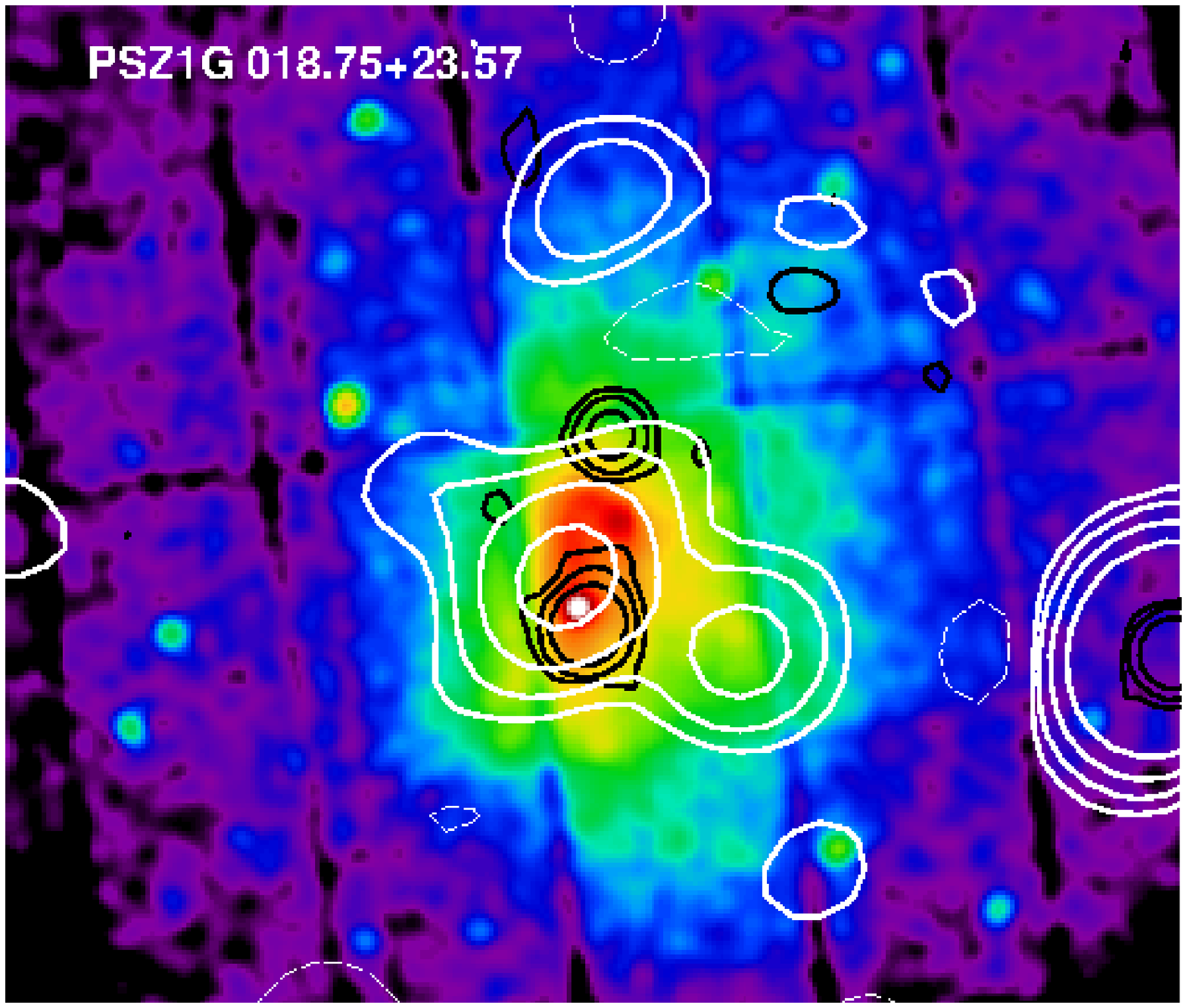}
\caption{KAT--7 residual contours (white) of the PSZ1G\,018.75+23.57 cluster after subtracting the compact sources identified in the NVSS images, overlaid on the {\it XMM--Newton} image. The contours are drawn at $\pm$~2.5, 5, 10, 20~mJy~beam$^{-1}$. The NVSS contours (black) are the same as Figure~\ref{fig:candidate}. The residual diffuse radio and X--ray emissions show a clear correlation.}
\label{fig:PSZ_residual}
\end{figure}

\begin{figure}
\centering
\includegraphics[width=1.2\columnwidth]{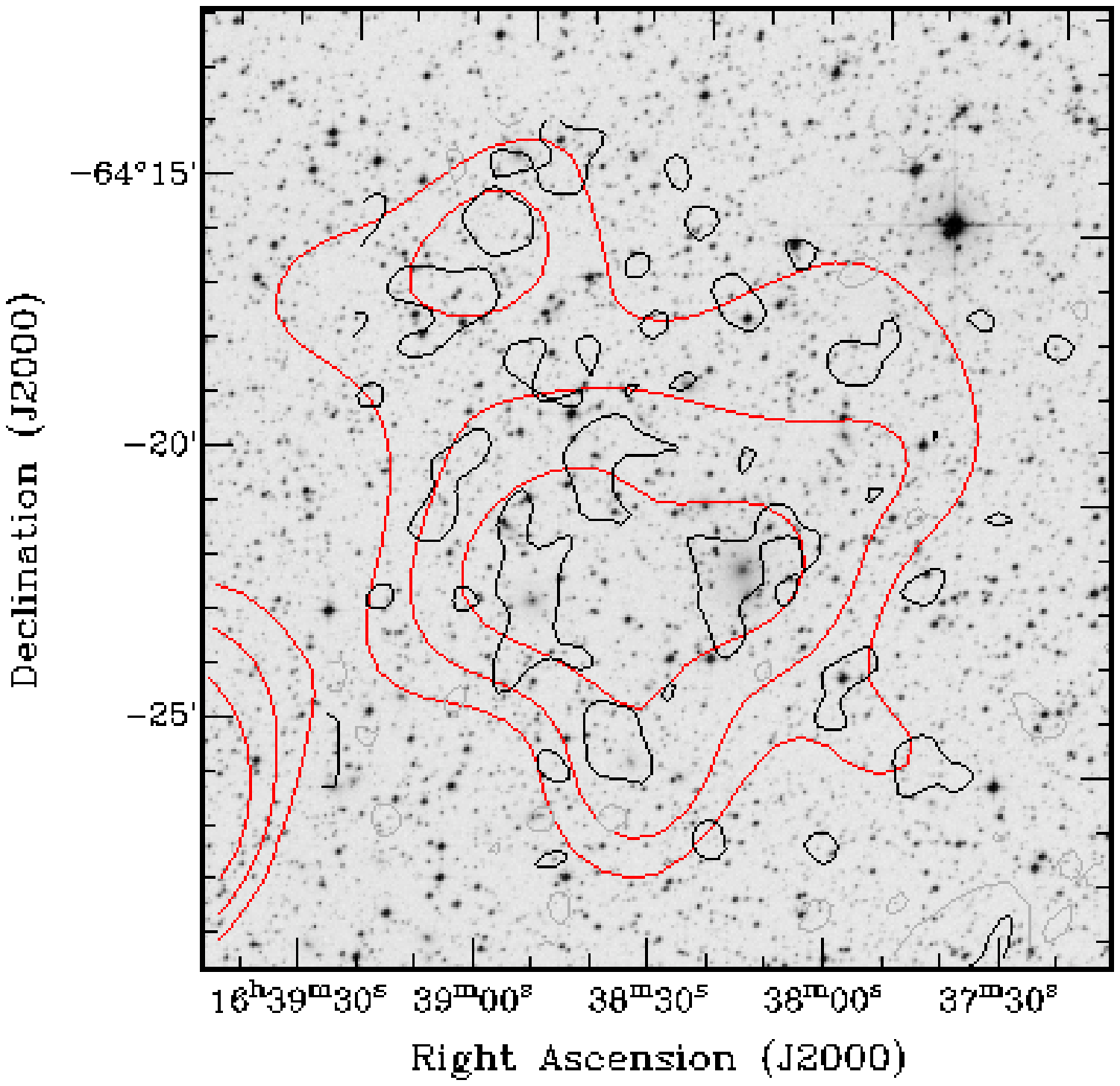}
\caption{1.86~GHz KAT--7 radio contours (red) of the Triangulum Australis cluster overlaid on the ESO Digitized Sky Survey optical image. The KAT--7 contours are the same as Figure~\ref{fig:candidate}, whereas the SUMSS contour (black) is drawn at 2~mJy~beam$^{-1}$ to highlight the match with bright optical galaxies.}
\label{fig:optical_Triangulum}
\end{figure}

\begin{itemize}

\item {\bf PSZ1G018.7+23.57}: Radio emission extends over more than 10~armin (Figure~\ref{fig:candidate}).
At the cluster position, 
the NVSS image shows two compact sources aligned north--south, following the same 
elongation of the cluster X--ray emission. The flux density of the two 
sources was extrapolated at 1.86~GHz assuming a fiducial spectral index 
$\alpha = 0.7$ and subtracted from the KAT--7 $uv$ data. The residual KAT--7 
image (Figure~\ref{fig:PSZ_residual}) shows a feature aligned East--West and coincident with the
X--ray emission, and a blob located north of this. The flux densities are 
S$_{\rm 1.86}=48.3 \pm 2.5$~mJy, and S$_{\rm 1.86}=9.0 \pm 0.5$~mJy
respectively. The extension of the central residual emission is $\sim 8$~arcmin, i.e. $\sim$~800~kpc. 
The $Newton-XMM$ image shows that the morphology of the cluster is highly 
disturbed.

\item {\bf A\,3628}: the radio emission is dominated by a strong source, 
S$_{\rm 1.86}=710\pm36$~mJy, with a fainter northern extension, 
which coincides with the peak of the X--ray emission. The 843~MHz radio 
contours from the SUMSS show that the northern extension is diffuse, with an 
extension of $\sim$~400~kpc. The $Chandra$ image clearly indicates that the 
cluster is dinamycally disturbed.

\item {\bf Triangulum Australis}: Scaife et al. (2015) first reported the presence of a radio halo at the 
centre of the Triangulum Australis cluster. In Figure~\ref{fig:candidate}
our image of the KAT--7 emission is overlaid on the $XMM-Newton$ image. 
The largest 
angular size of the diffuse emission is $\sim 15$~arcmin, corresponding to
$\sim 900$~kpc. Its morphology is double--peaked at the centre, consistent
with Scaife et al. (2015), however the position angle of the emission
shown here is slightly different. An overlay between our KAT--7 image
and the red optical frame from the ESO Digitized Sky Survey (DSS--2, Figure~\ref{fig:optical_Triangulum}) shows
that our two peaks are coincident with two very bright galaxies (the
western one most likely being the cluster brightest dominant galaxy), so the
inner double morphology may be contaminated by individual emission, calling for higher angular resolution observations to 
isolate the contribution from individual galaxies.
The emission detected at 843~MHz with SUMSS is patchy, as a result of
inadequate coverage at the largest angular scales, and it is spread 
all over the KAT--7 emission. 
We measure a S$_{\rm 1.33}=92 \pm 5$~mJy flux density for the KAT--7 diffuse emission that, albeit lower, is still compatible at the 3$\sigma$ confidence level with the value reported by Scaife et al. (2015). 

\end{itemize}
%
%
%

\subsection{Limits to the radio power of undetected giant RHs}
\label{sec:upper_limits}

With the aim to set an upper limit to the radio power in those clusters void of radio emission at their centres, we followed the approach presented in Venturi et al. (2008) and Kale et al. (2013), which we refer to as ``injections''. In particular:

1) the typical radio brightness distribution of a Mpc--sized RH was modelled on the basis of well studied RHs (Cassano et al. 2007, Brunetti et al. 2007). The typical model we adopted is made of seven concentric spheres with increasing radius (from 43 to 720~arcsec) and increasing flux density (i.e. the largest components are also those with the highest flux density fraction). The angular size of 720~arcsec corresponds to a liner size of 1~Mpc at z~$= 0.074$. In order to appropriately span the redshift range of our cluster sample, the sizes were multiplied by a factor of either 0.8 or 1.2. The brightness profiles $I(r)$ of our injected RH models as a function of radius $r$ are consistent with an exponential profile $I(r) \propto e^{-\frac{r}{r_s}}$ with $120 \leq r_s \leq 170$~kpc. This choice of $r_s$ is consistent with the e-folding radii measured in tyipical RHs (Murgia et al. 2009).

We explored the $50 < {\rm S}_{inj} < 1000$~mJy range of total flux density for each model. The surface brightness of the various components in the same model ranges from $\sim$~3$ \times 10^{-5}$ to $\sim$~1.5$ \times 10^{-4}$~mJy~arcsec$^{-2}$, referred to a halo with a total flux density of 50~mJy.

2) each modelled brightness distribution was ``injected'' in the $uv$ data ($uv_{inj}$) using the UVMOD task in AIPS (Astronomical Imaging Processing System);

3) each dataset $uv_{inj}$ was imaged and we measured the flux density integrated over the area corresponding to the Mpc region (S$_{rec}$, recovered flux density), and compared it to the total injected flux density, S$_{inj}$.

In order to evaluate how the method of the injections performs with KAT--7 data, we chose an empty central area in two clusters with different $uv$~coverage due to declination, namely RXCJ\,1358.9--4750 and A\,644 that is located at DEC$_{\rm J2000} \sim -7^\circ$ (see Figure~\ref{fig:uv-coverage}).

As expected, S$_{rec}$/S$_{inj}$ increases with increasing S$_{inj}$, going from $\sim$~$7\%-12\%$ for S$_{inj}$~=~50~mJy to
$\sim$~$40\%-60\%$ for S$_{inj}$~=~300~mJy, up to $\sim$~80\% for S$_{inj}$~=~1~Jy, the lower limits referring to those cases with both lower surface brightness and worse $uv$~coverage (i.e A\,644).

Based on those results we conclude that a RH should have a minimum flux density S$^{min}_{inj}$~=~50~mJy in order to be detectable in the KAT--7 images. For lower S$_{inj}$ values, the central part of the resulting image appears just to be compatible with noise fluctuations, while for S$_{inj}$~=~50~mJy we measure a residual flux density of the order of $\sim$~$4-10$~mJy, which is $2-5$ times the average integrated flux density at the centre of those clusters with no radio emission.

We applied our analysis to seven clusters (see Table~\ref{tab:obs}), i.e. the five clusters void of emission at their centres (see Section~\ref{sec:noradio}) plus A\,3158 and A\,3921, as we could reliably subtract the point source contribution from the data (see Section~\ref{sec:scrs}). The RH upper limit was scaled from 1.86~GHz to 1.4~GHz using a spectral index $\alpha$~=~1.3.

%
\begin{figure}
\centering
\includegraphics[width=0.9\columnwidth]{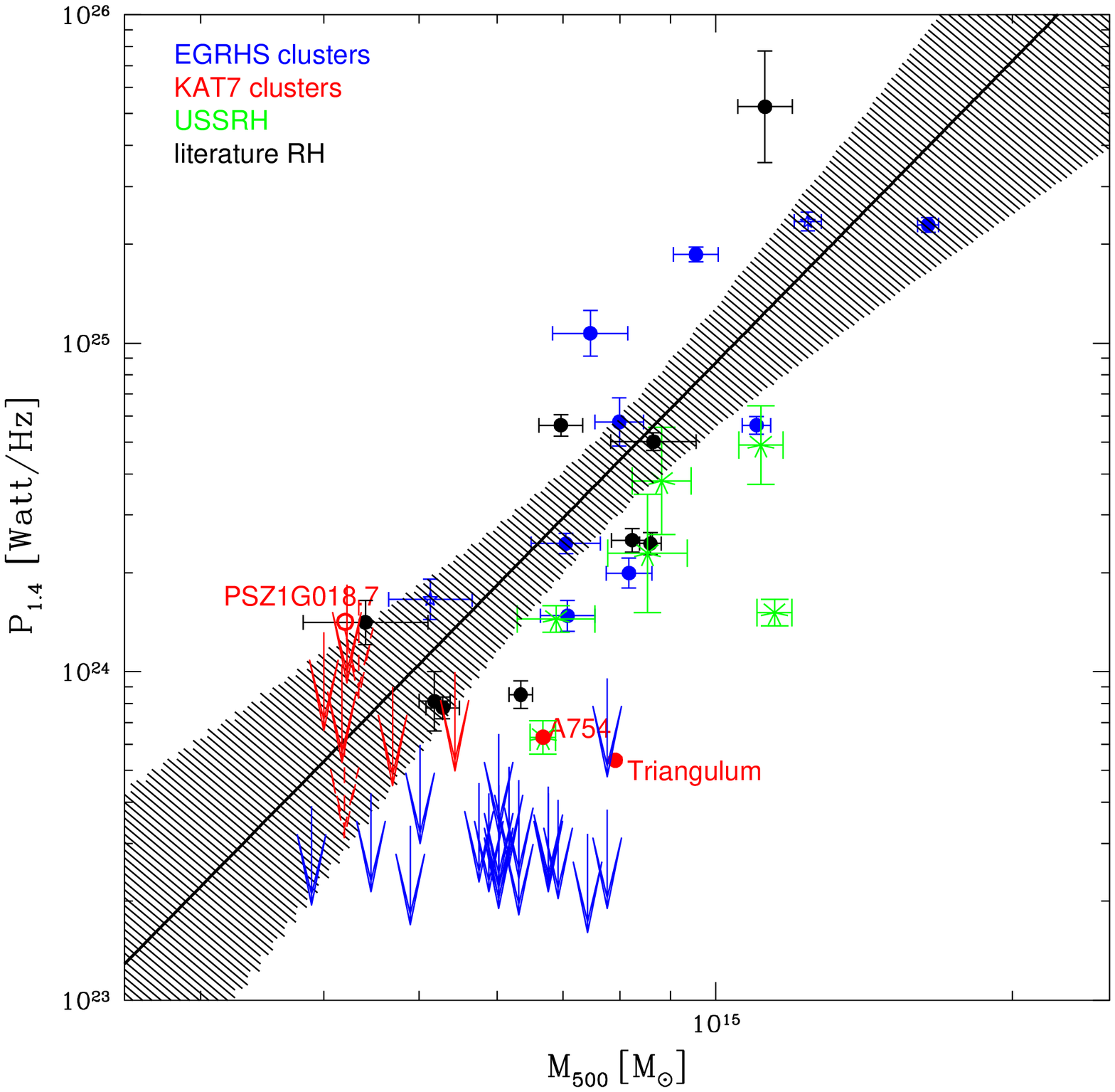}
\caption{Distribution of clusters in the $P_{\rm 1.4}-M_{500}$ diagram. 
Different symbols indicate RHs belonging to the EGRHS (blue dots), RHs 
from the literature (black dots), RHs with very steep spectra (USSRH, 
green asterisks); upper limits from the EGRHS (blue arrows) and upper limits 
derived in the present paper with KAT--7 (red arrows). 
The best-fit relation to giant RHs only (black line) is shown together with its 95\% confidence region (shadowed region).}
\label{fig:LrM500}
\end{figure}

\subsection{The radio halo power vs cluster mass correlation}
								    
A clear correlation between the synchrotron power of RHs at 1.4~GHz, $P_{\rm 1.4}$, and the mass of the hosting clusters, $M_{500}$, was found by Basu et al. (2012) through the analysis of the early Planck catalogue.
Cassano et al. (2013), using EGRHS data and the Planck SZ cluster catalogue (Planck Collaboration 2014), confirmed this relation and found a bimodal behaviour in the $P_{\rm 1.4}-M_{500}$ diagram for clusters with $M_{500}\simgt 5 \times 10^{14}\,M_{\odot}$ (similar to, although weaker than, that found in the radio/X-ray plane; {\it e.g.} Brunetti et al. 2007, 2009), where clusters with RHs follow a steep relation 
$P_{\rm 1.4}\propto M_{500}^{\sim 3-4}$ and clusters without RHs are all well below the 95\% confidence region of the best-fit correlation. This is clear from Figure~\ref{fig:LrM500}, that shows the $P_{\rm 1.4}-M_{500}$ correlation obtained using well studied giant RHs in the literature (see Table~1 in Cassano et al. 2013), together with upper limits on the radio power of clusters without giant RHs derived for the EGRHS sample (Venturi et al. 2007, 2008; Kale et al. 2013, 2015).

We used the KAT--7 results to extend this study to less massive clusters.
A\,754 and the Triangulum Australis (shown and labelled in red in Figure~\ref{fig:LrM500}) are the only two clusters in our sample with a RH reported in the literature. The Triangulum Australis RH lies significantly below the correlation and its flux density measurement may still be affected by the unsubtracted compact sources as well as by limitations in the $uv$ coverage as discussed in Section~\ref{sec:candidate_radio_halos} and~\ref{sec:upper_limits} respectively. 
Based on the results of the injection simulations and the magnitude of the measured RH flux density, it is unlikely that a large fraction of its real flux density is missing in our observations. We thus conclude that this RH lies indeed below the correlation, 
in the area where ultra steep spectrum RHs are found. The candidate RH in PSZ1G\,018.75+23.57 lies instead within the 95\% confidence region of the correlation.

All KAT--7 upper limits lie within the 95\% confidence region of the best-fit correlation.
This result does not allow to constrain a bi-modal distribution, but it statistically rules out the presence of powerful RHs in smaller mass systems ($M_{500}\simlt 5 \times 10^{14}\,M_{\odot}$).

\section{Discussion and conclusions}
\label{final_conclusions}

We have presented KAT--7 observations at 1.86~GHz of a mass-selected 
(M~$ > 4 \times 10^{14}$~M$_{\odot}$) sample of 14 galaxy clusters aimed at 
studying the occurrence of giant RHs. This sample was completed by the analysis of KAT--7 archive data for the Triangulum Australis cluster.

Three clusters host a candidate RH, as suggested by the comparison
between our KAT--7 images, NVSS and SUMSS images and the X--ray emission from
the hot intracluster gas. These are: the Triangulum Australis, where 
the presence of an almost Mpc size radio diffuse has been recently
reported by Scaife et al. (2015), PSZ1G\,018.75+23.57 and A\,3628.

Seven clusters in the sample host emission at their centre. For A\,3158 and
A\,3921 the combination of our images with radio information at higher 
angular resolution available in the literature suggests that the
KAT--7 emission is from compact sources. For RXCJ1407.8-5100 and 
RXCJ2201.9-5956 the comparison between the KAT--7 and the SUMSS emission 
suggests that our 1.86~GHz radio images are most likely a blend of 
individual sources. For the remaining three clusters, i.e. A\,3266, 
A\,3695 and A\,3911, our limited angular resolution and the lack of adequate
high resolution imaging do not allow an unambiguous classification.

For seven clusters in our sample we could set an upper limit to the presence 
of a giant RH, which we report on the 
$P_{\rm 1.4}-M_{500}$ correlation together with the flux density measured for the candidate RHs in PSZ1G\,018.75+23.57 and in the Triangulum Australis. As A\,85 was the only cluster with available halo information in the $4 < {\rm M} < 5.5 \times 10^{14}$~M$_{\odot}$, our data offer the first statistical information about cluster RHs in this mass range and confirm the lack of bright RHs in smaller systems, indicating that RHs more powerful than expected from the correlation must be rare. Under the assumption that the $P_{\rm 1.4}-M_{500}$ correlation can be extended to smaller systems, our results indicate that this correlation should show a steep slope, of the form $P_{1.4}\propto M_{500}^{\beta}$, with $\beta \simgt 3$.

\section*{Acknowledgments}
We thank the referee for useful comments that helped improving the manuscript. This work is based on research supported by the National Research Foundation under grant 92725. Any opinion, finding and conclusion or recommendation expressed in this material is that of the author(s) and the NRF does not accept any liability in this regard. This work was also partly supported by the Executive Programme of Scientific and Technological Co-operation between the Italian Republic and the Republic of South Africa 2014--2016. The KAT--7 is supported by SKA South Africa and by the National Science Foundation of South Africa.

\appendix
\section{KAT--7 cluster images}
\label{appendix_A}

We display the $2^\circ \times 2^\circ$, full primary beam KAT--7 image for the all the clusters in our sample. Images have not been corrected for the primary beam attenuation. Residual sidelobes around the strongest sources are likely related to residual errors in the amplitude calibration, as phase--only self--calibration has been applied (see Section~\ref{sec:obs}).

\begin{figure}
\centering
\includegraphics[width=0.8\columnwidth,angle=270]{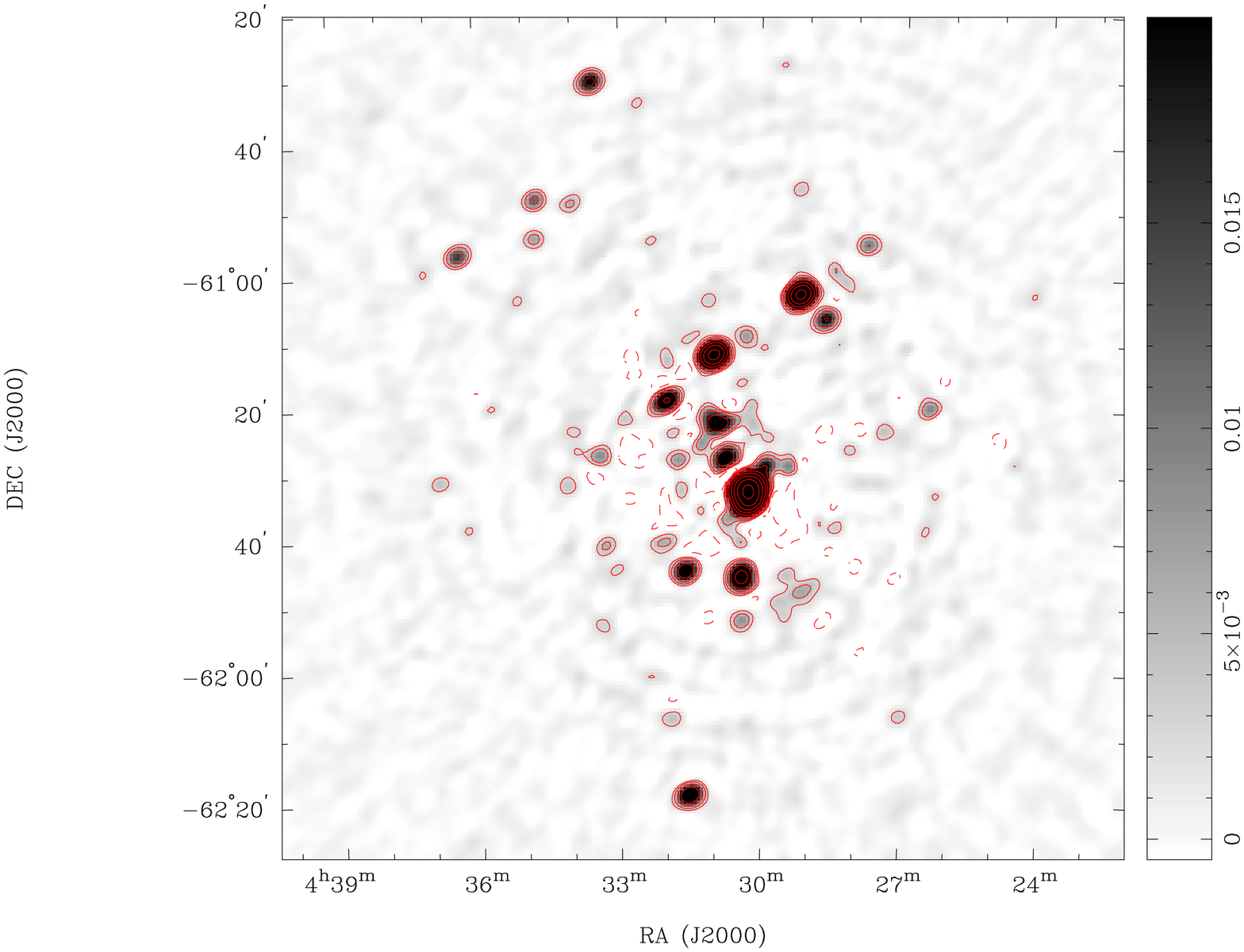}
\includegraphics[width=0.8\columnwidth,angle=270]{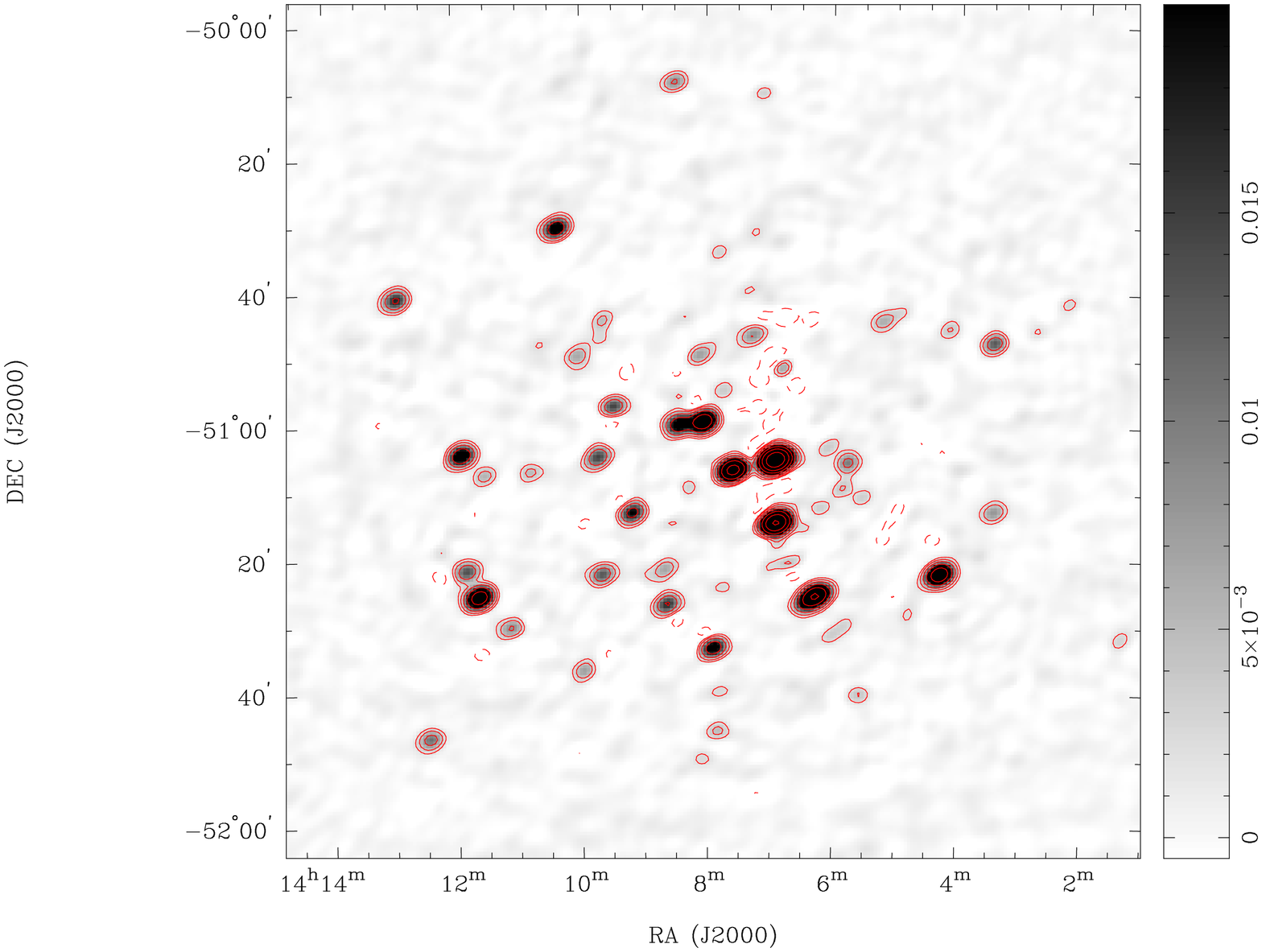}
\includegraphics[width=0.8\columnwidth,angle=270]{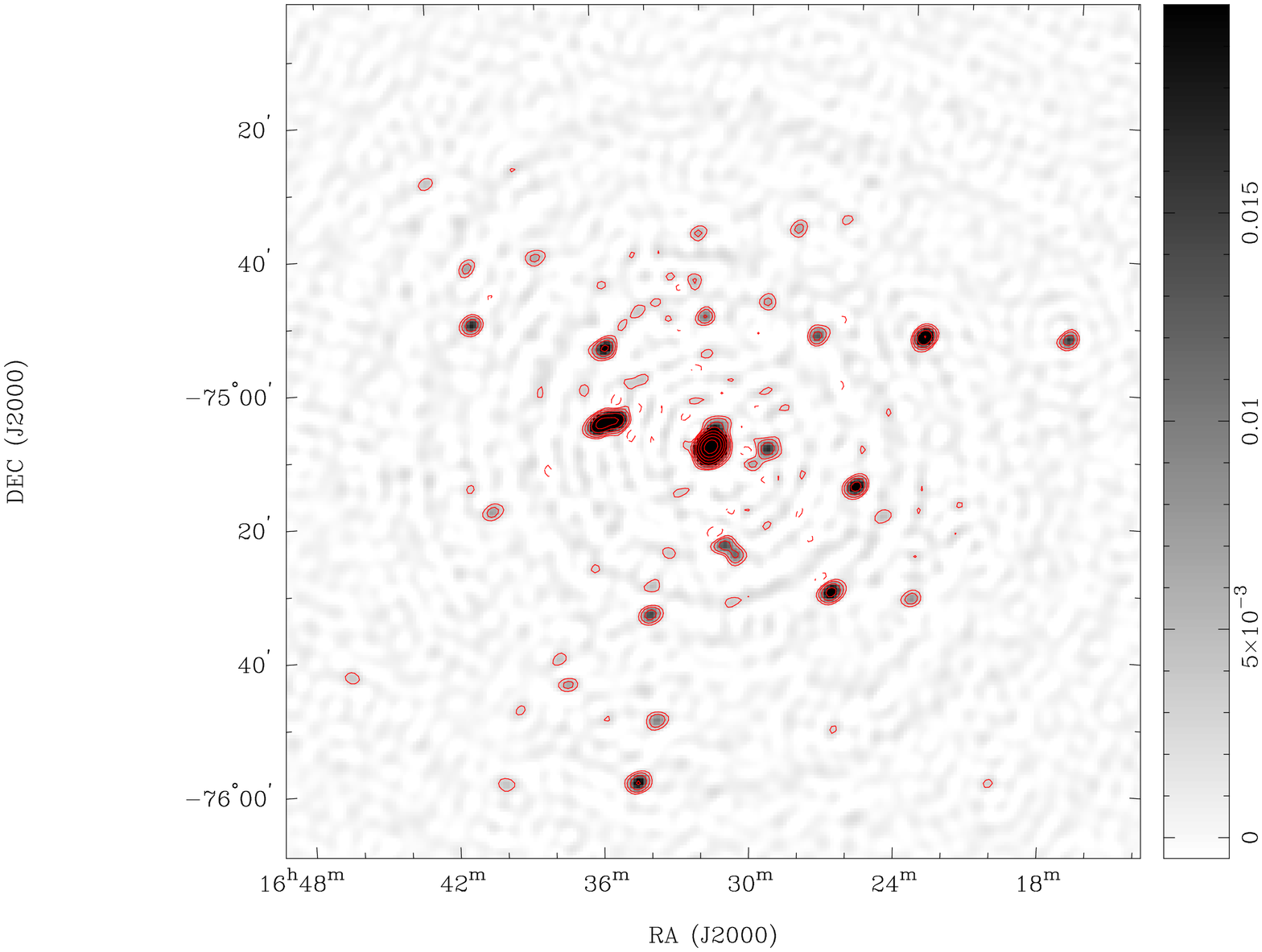}
\caption{KAT--7 images for A3266 (top), RXCJ1407 (middle) and A3628 (bottom). The color scale is in Jy~beam$^{-1}$ and the contours are drawn at -3, 3, 6, 12, 24, 48... $\times$ the rms noise values reported in Table~\ref{tab:obs}.}
\label{fig:app1}
\end{figure}
\begin{figure}
\centering
\includegraphics[width=0.8\columnwidth,angle=270]{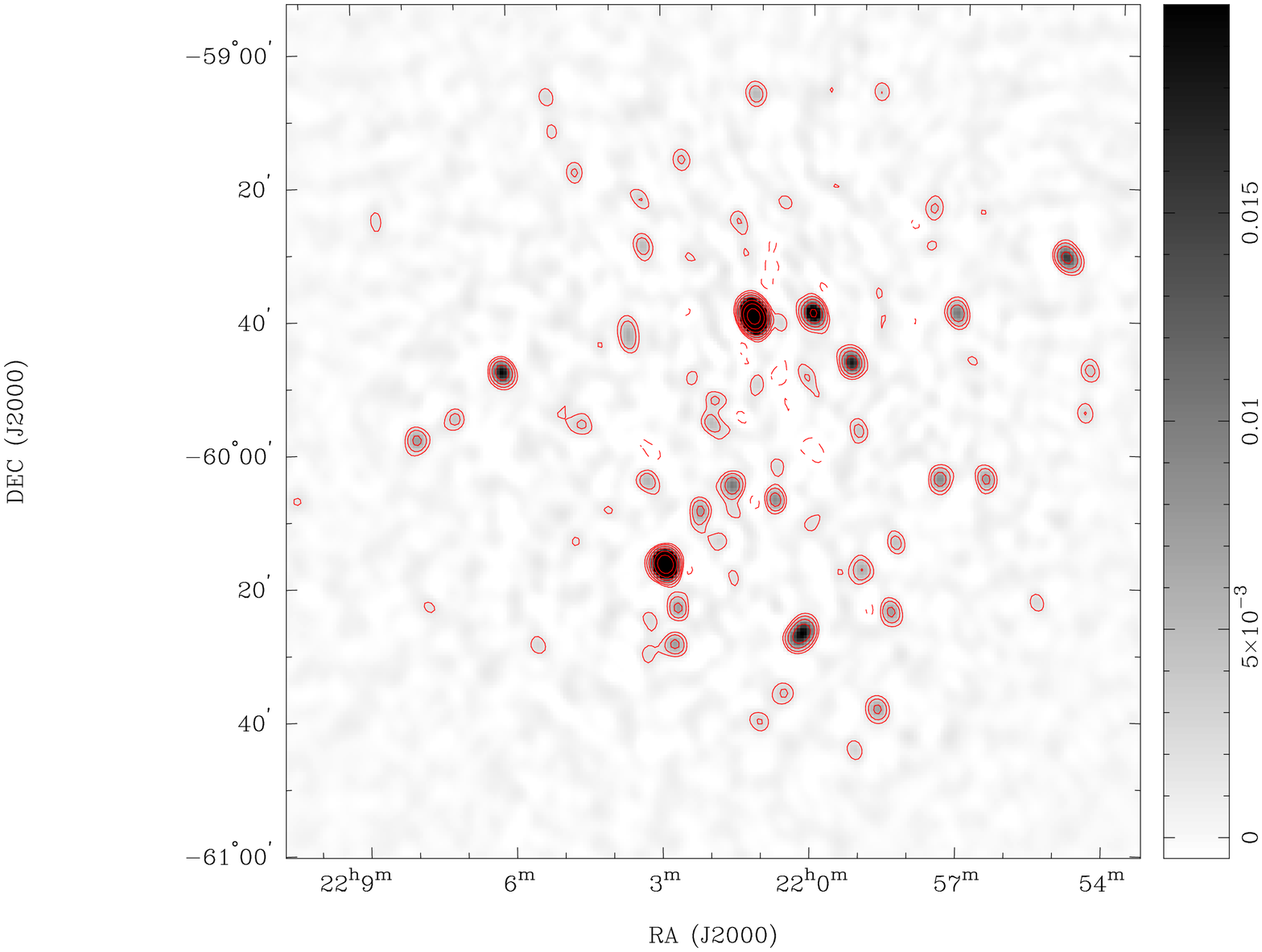}
\includegraphics[width=0.8\columnwidth,angle=270]{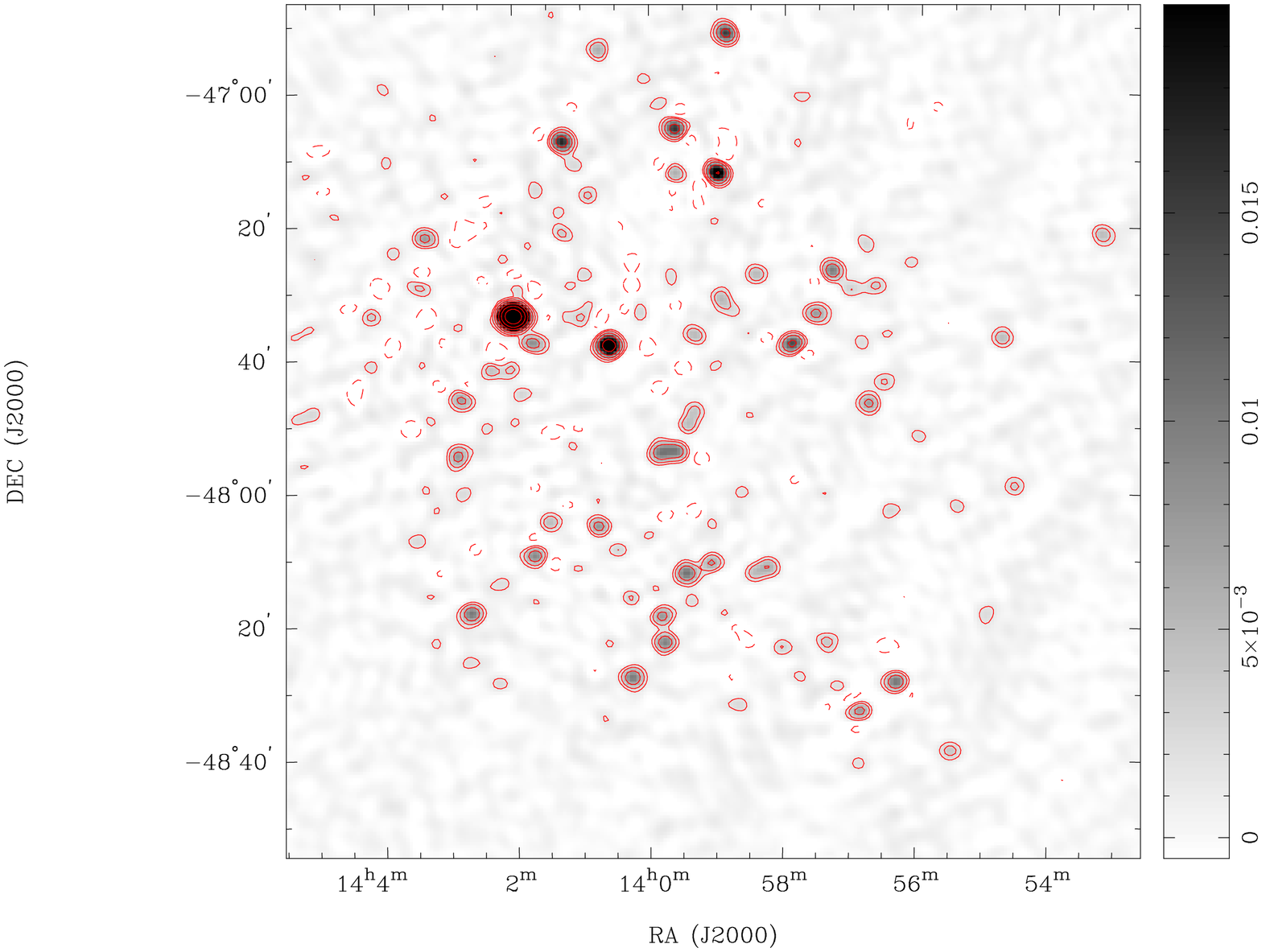}
\includegraphics[width=0.8\columnwidth,angle=270]{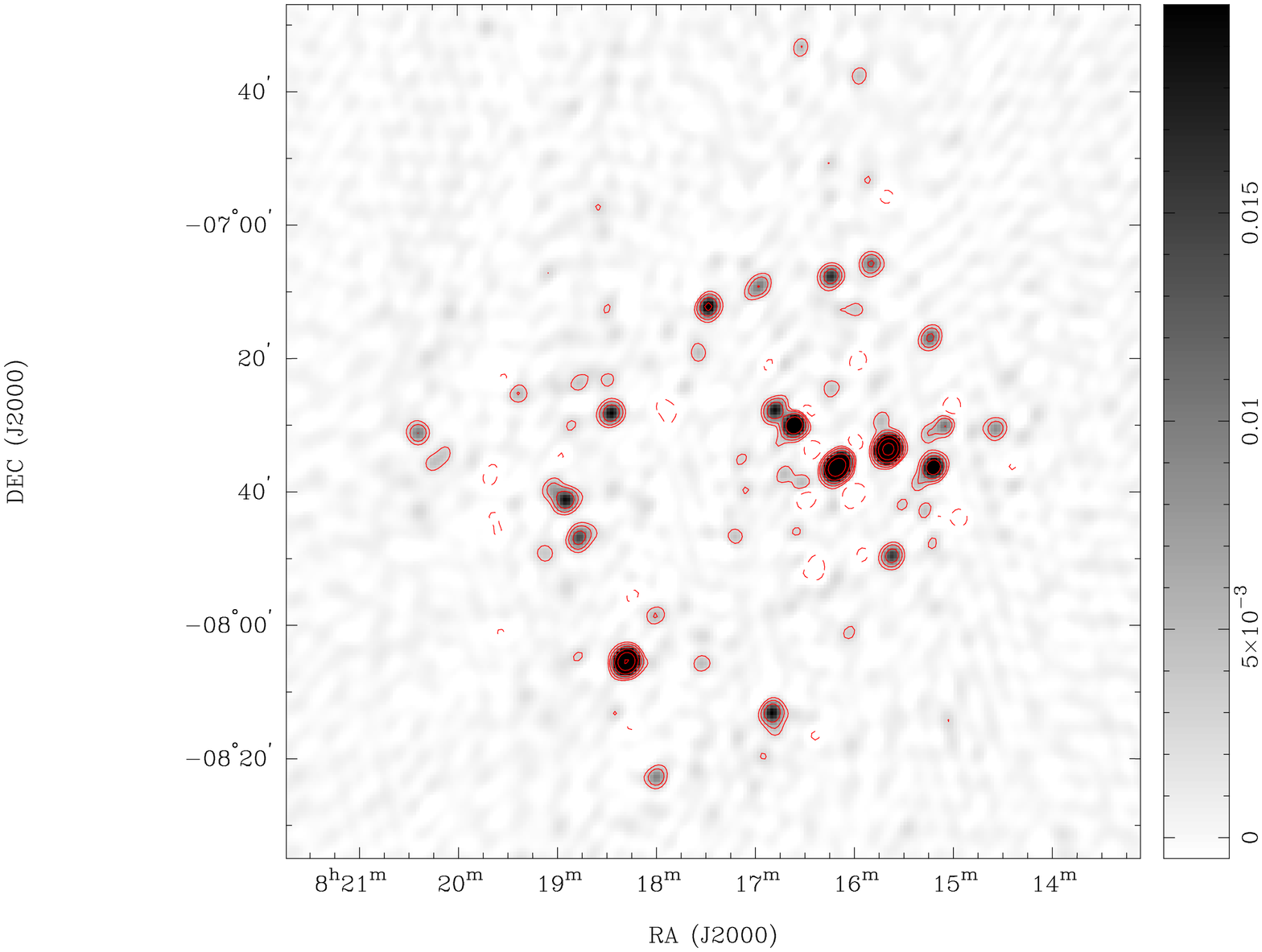}
\caption{Same as Figure~\ref{fig:app1} but for A\,3827 (top), RXCJ1358 (middle) and A644 (bottom).}
\label{fig:app2}
\end{figure}
\begin{figure}
\centering
\includegraphics[width=0.8\columnwidth,angle=270]{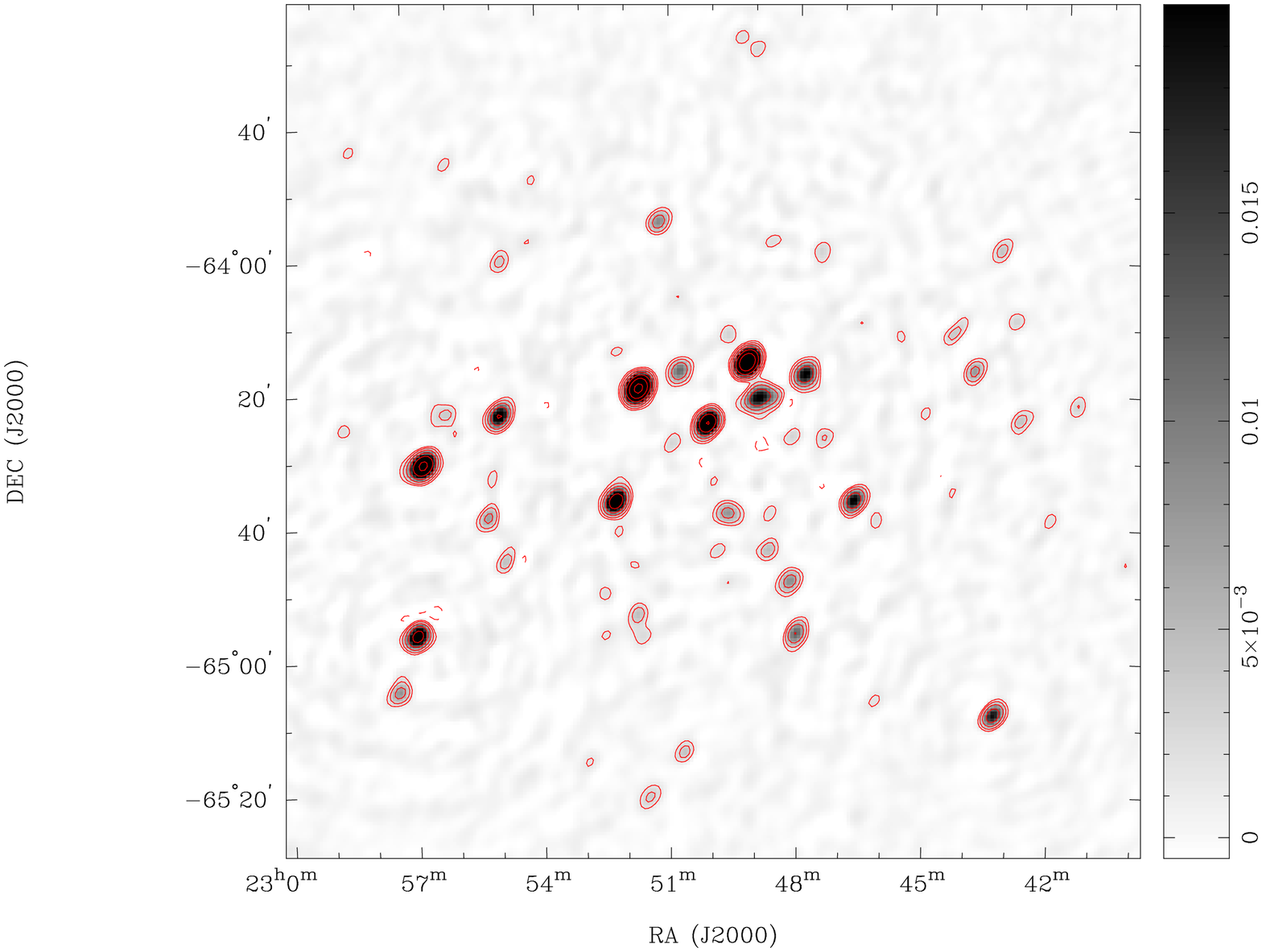}
\includegraphics[width=0.8\columnwidth,angle=270]{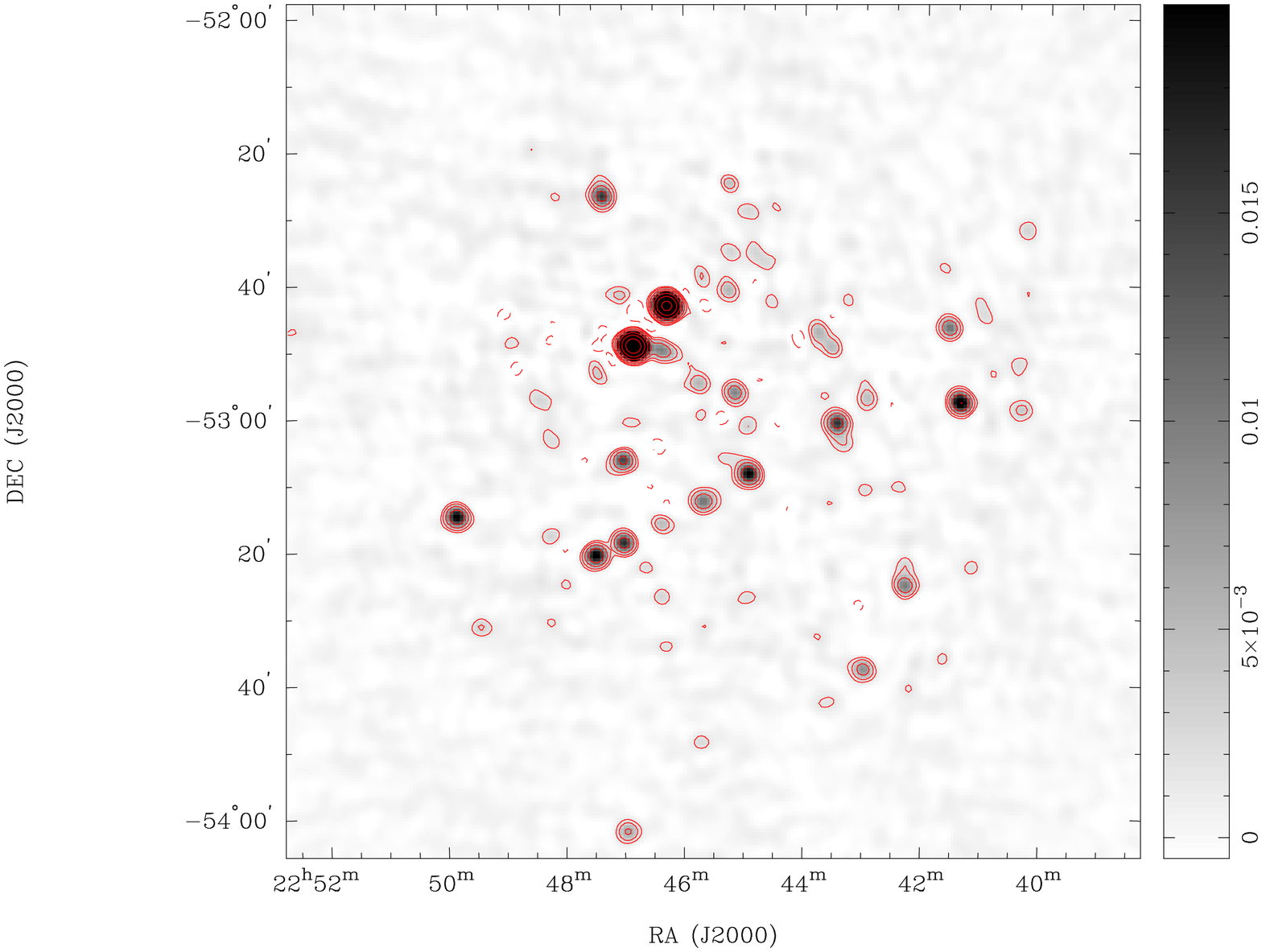}
\includegraphics[width=0.8\columnwidth,angle=270]{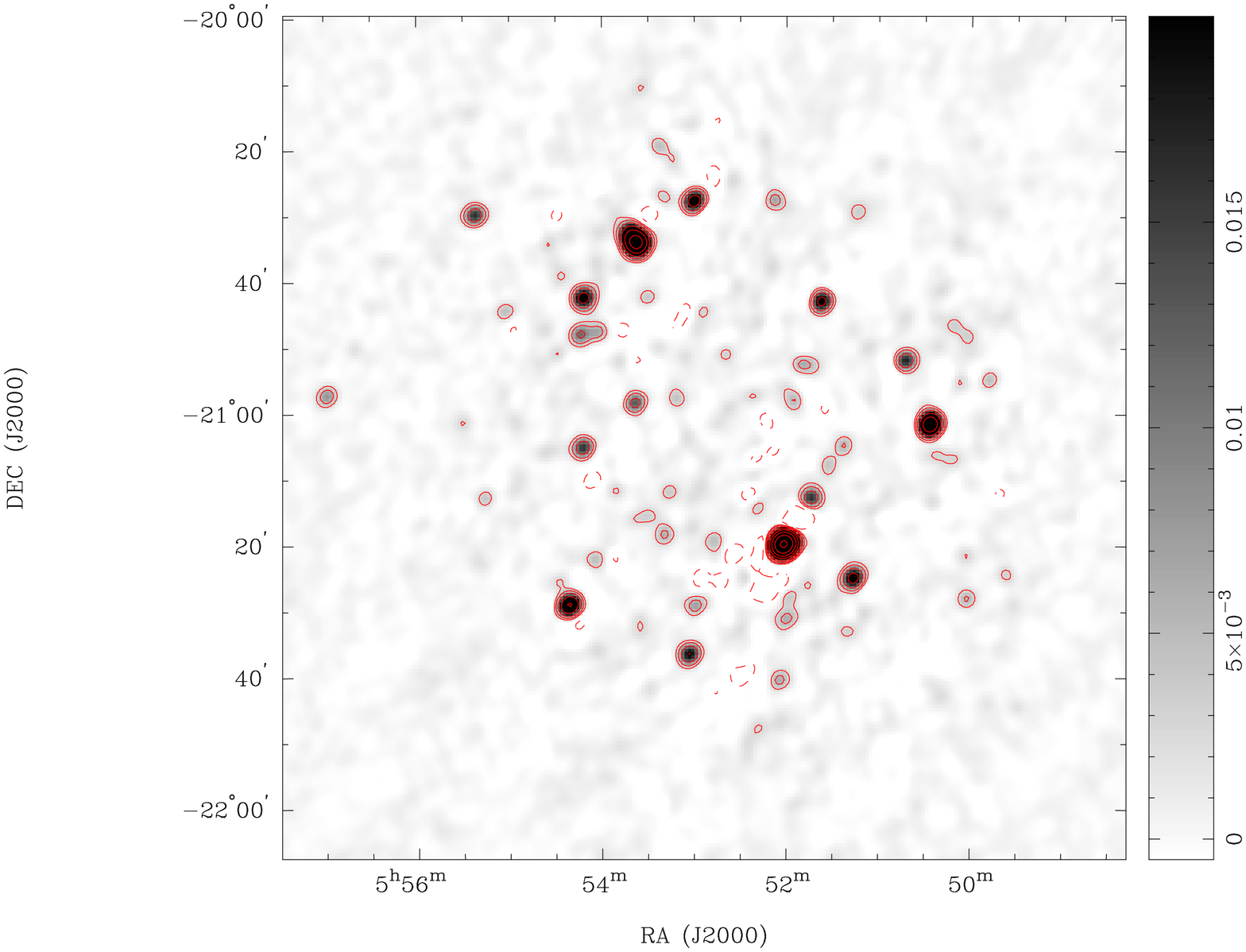}
\caption{Same as Figure~\ref{fig:app1} but for A3921 (top), A3911 (middle) and A550 (bottom).}
\label{fig:app3}
\end{figure}
\begin{figure}
\centering
\includegraphics[width=0.8\columnwidth,angle=270]{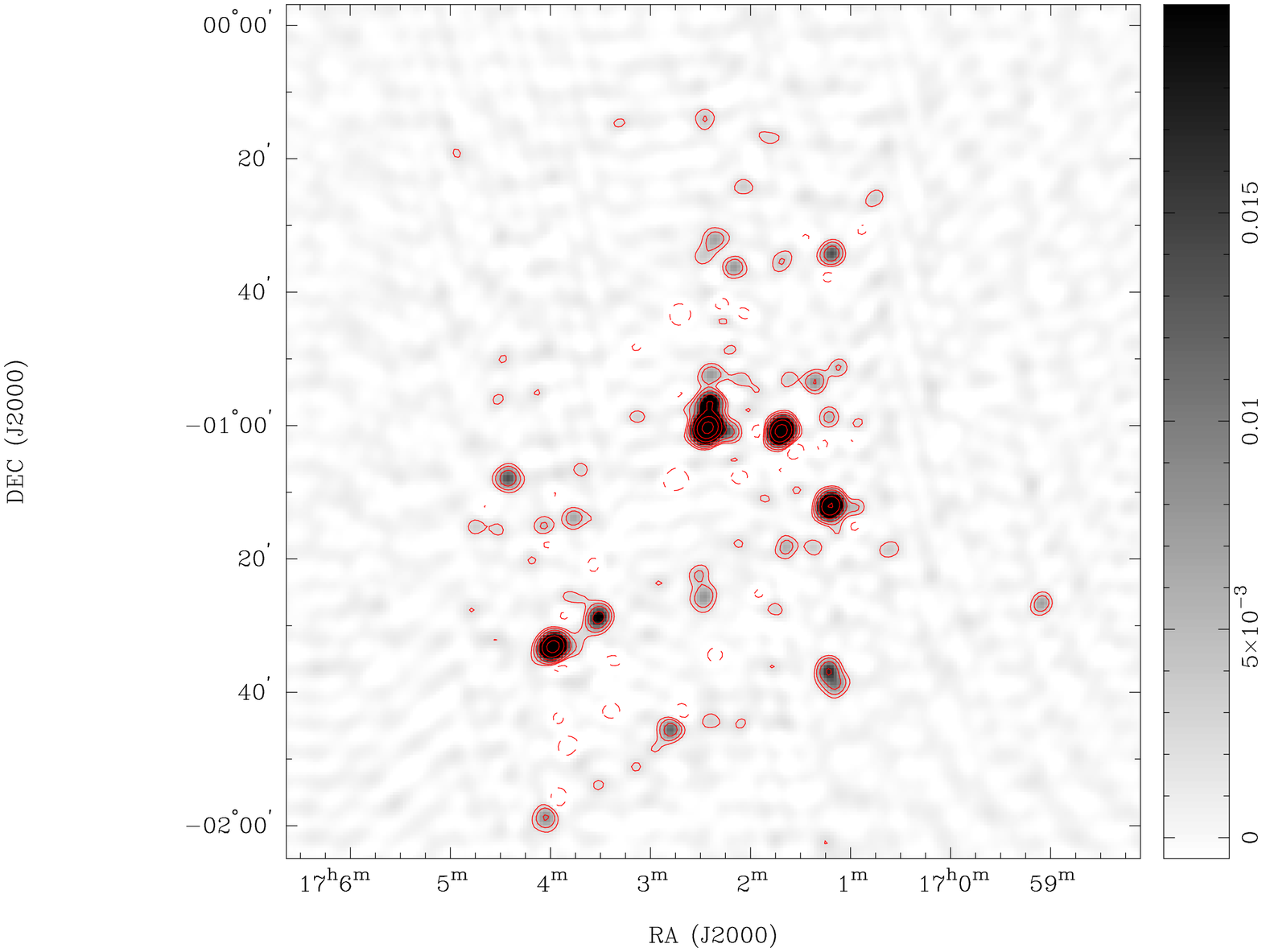}
\includegraphics[width=0.8\columnwidth,angle=270]{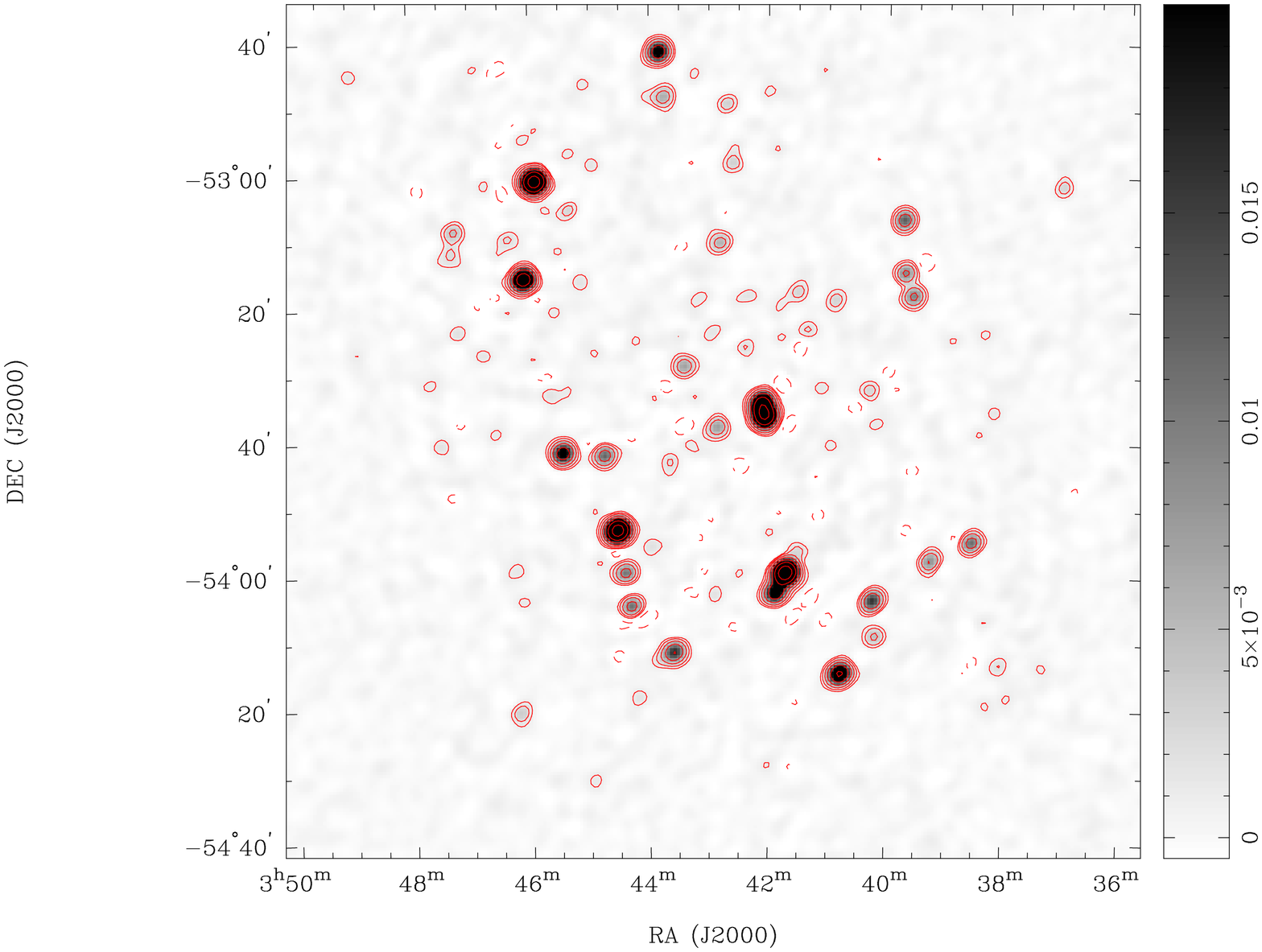}
\includegraphics[width=0.8\columnwidth,angle=270]{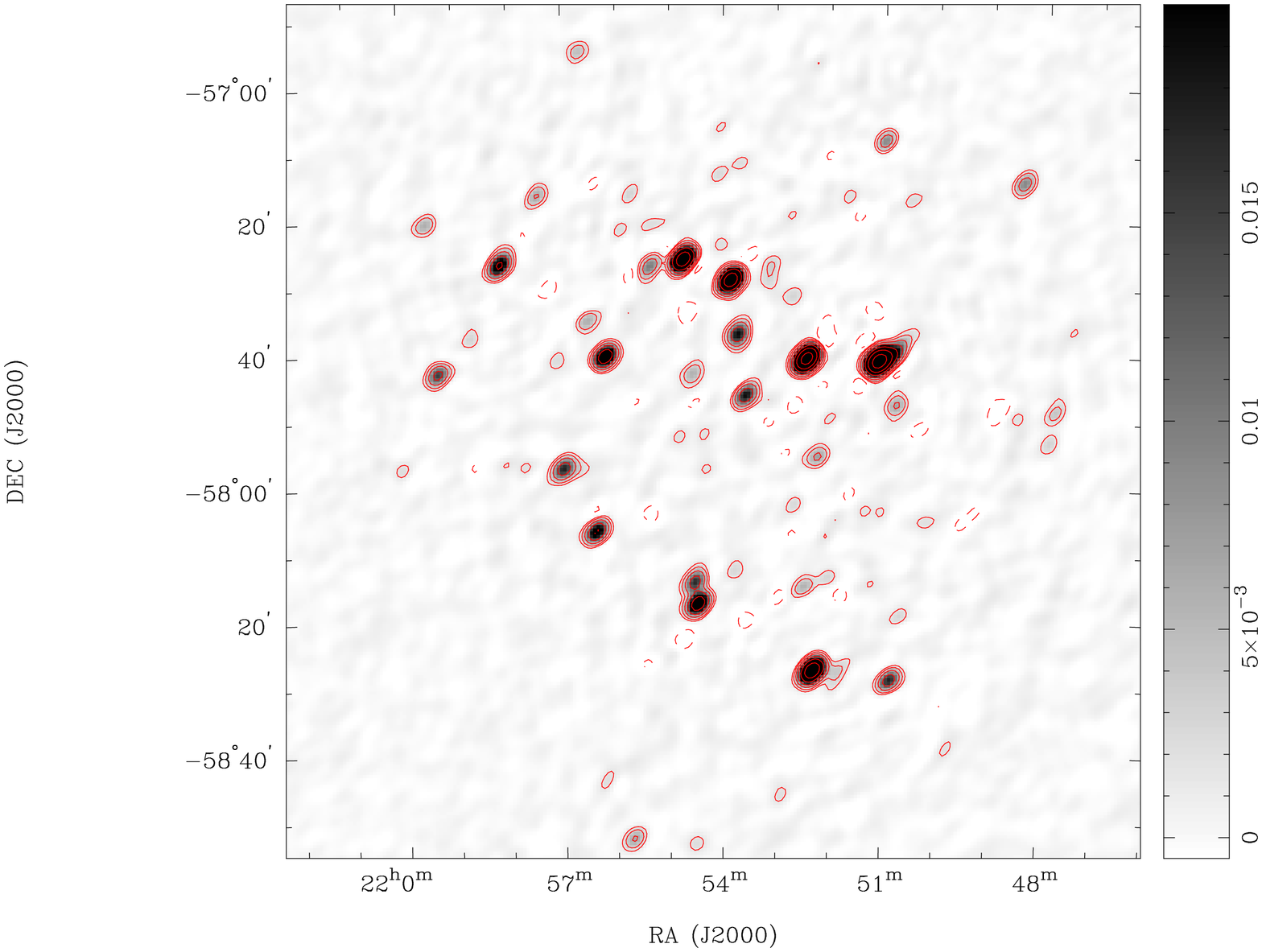}
\caption{Same as Figure~\ref{fig:app1} but for PSZ1G018 (top), A3158 (middle) and A3822 (bottom).}
\label{fig:app4}
\end{figure}
\begin{figure}
\centering
\includegraphics[width=0.8\columnwidth,angle=270]{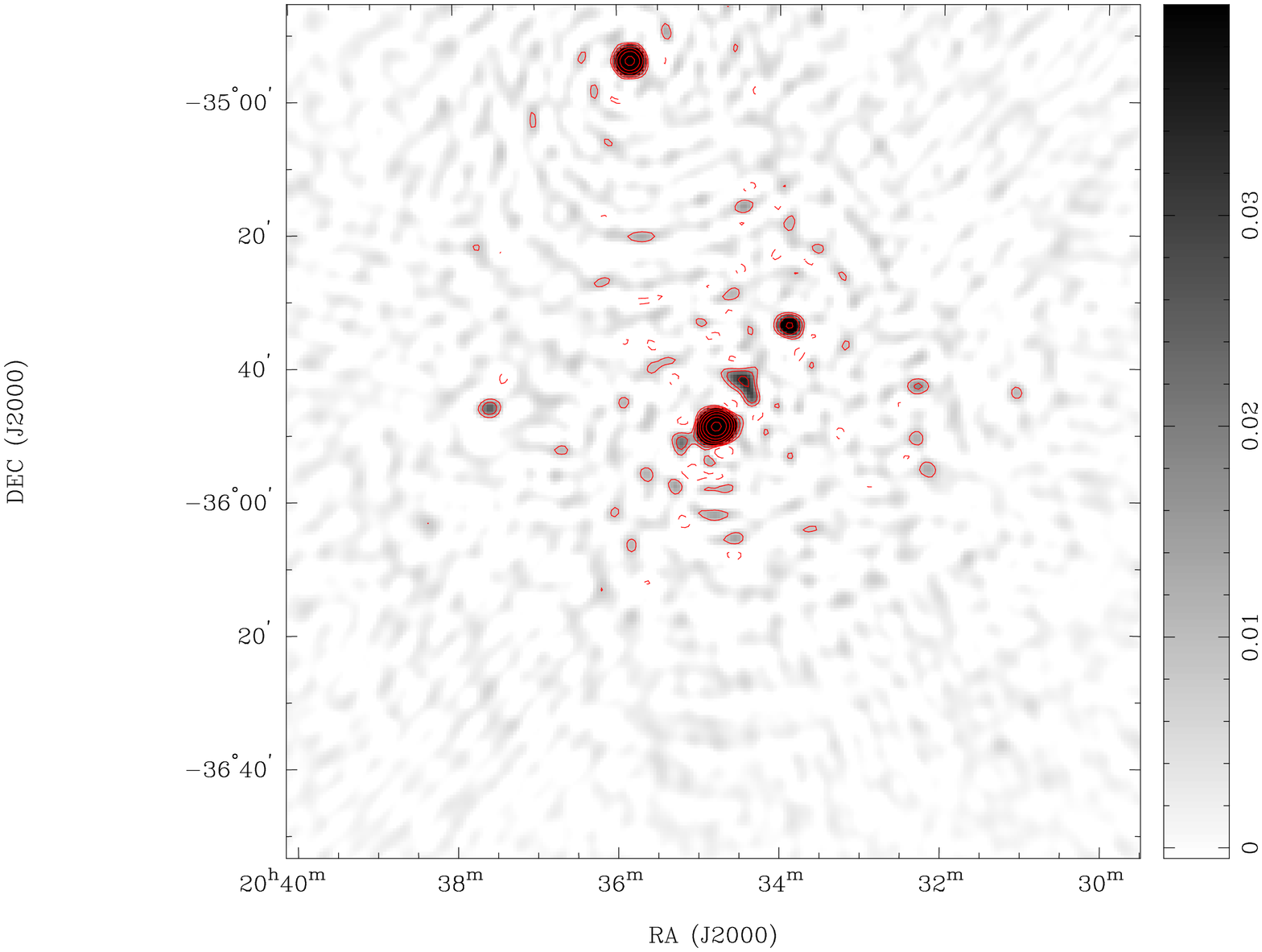}
\includegraphics[width=0.8\columnwidth,angle=270]{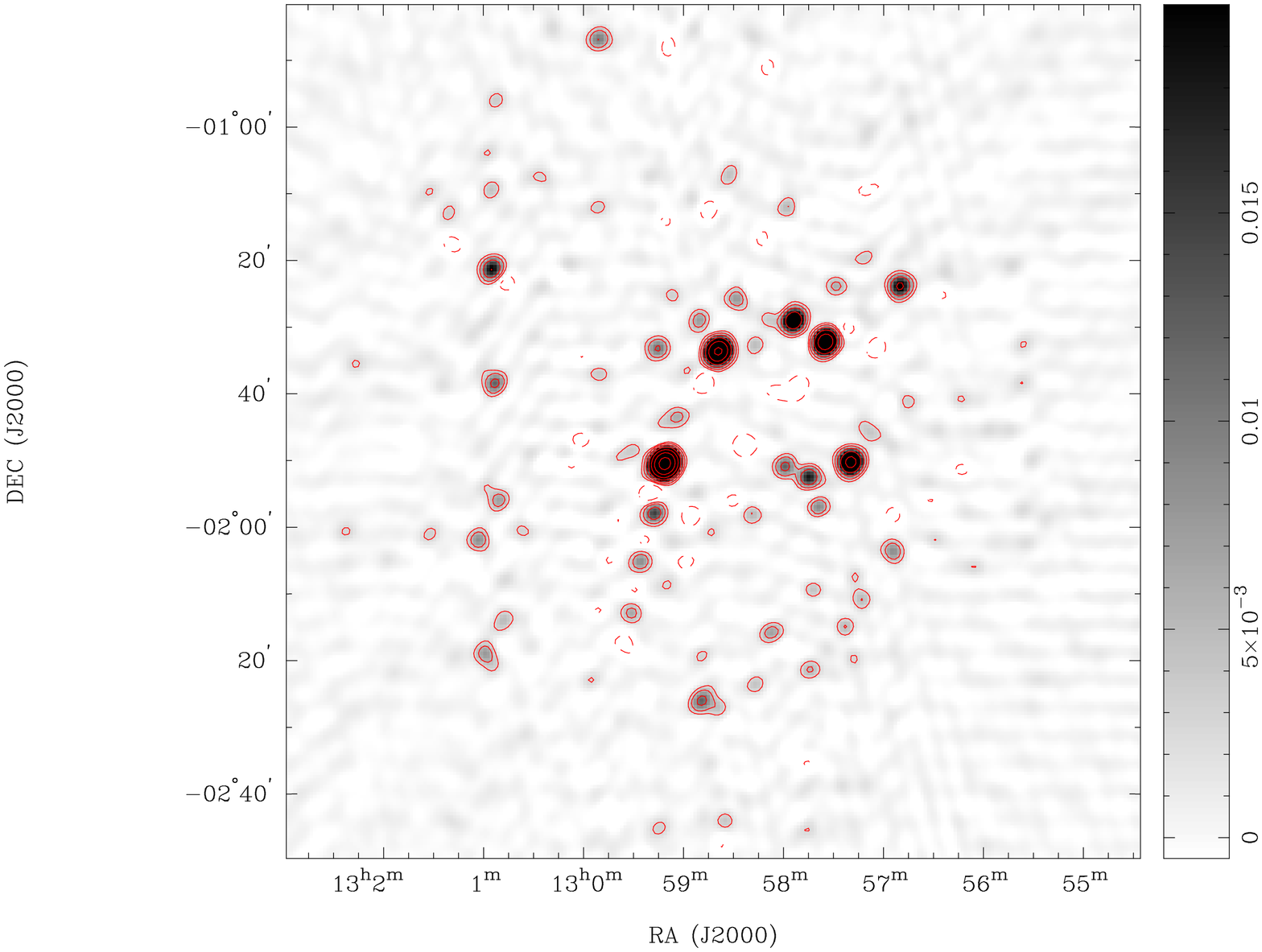}
\caption{Same as Figure~\ref{fig:app1} but for A3695 (top) and A1650 (bottom).}
\label{fig:app5}
\end{figure}

\label{lastpage}

\end{document}